\def\alt{\hbox{\raise.5ex\hbox{$<$}
\kern-1.1em\lower.5ex\hbox{$\sim$}}}

\documentstyle[]{l-school}

\begin{document}
\markboth{L. Blanchet}{HOUCHES}

\setcounter{part}{1}

\title{Gravitational radiation \\ from relativistic sources}
\author{Luc Blanchet}
\institute{D\'epartement d'Astrophysique Relativiste et de Cosmologie,\\
Centre National de la Recherche Scientifique (UPR 176),\\
Observatoire de Paris, 92195 Meudon Cedex, France}

\maketitle

\section{INTRODUCTION}  \label{sec:1}

Albert Einstein \cite{E16} found some radiative solutions to
the field equations of general relativity and deduced
the existence of gravitational waves, which could be interpreted as ripples
in the curvature of spacetime propagating with the speed of light.  In a
subsequent paper \cite{E18}, he computed the total power emitted in the form
of gravitational waves by an isolated ``Newtonian'' source, and found that
this power depends quadratically on the variations of the quadrupole moment of
the source in a celebrated formula known today as Einstein's quadrupole
formula.  The fact that gravitational waves from an isolated source are
dominantly quadrupolar is a consequence of the equivalence principle, which
implies the conservation (or linear variation in time) of the source's
monopole and dipole moments as the consequence of the equations of motion.

Existence of gravitational radiation was confirmed by Eddington \cite{Edd23}
who showed that the propagation at the speed of light of weak gravitational
waves on a flat background has an intrinsic meaning (independent of the
choice of a coordinate system), and by Bondi \cite{B57} who proved that it is
possible to construct a detector of gravitational waves which is heated when
the wave passes through it. Surely gravitational radiation thus carries energy.
Furthermore this energy is extracted from the mass-energy of the source
(as measured from infinity), which has been shown by Bonnor \cite{Bo59} and
Bondi, van der Burg and Metzner \cite{BBM62} (see also Sachs \cite{S62}) to be
a monotonically decreasing function of time during the emission. Before
emission this mass-energy reduces to the mass of Arnowitt, Deser and Misner
\cite{ADM} which represents the total mass-energy of the source plus the
contribution of the waves to be emitted. Some theorems due notably to Shoen
and Yau \cite{SY79,SY82} have established the positivity of both the ADM and
Bondi mass of an isolated system. This proves in particular that the mass
of the source bounds from above the total amount of energy which it can
emit in the form of radiation.  All these theoretical works show that
the classic vision of a wave carrying energy, which is extracted from
its source and can be put down an antenna, is legitimate in the case of
gravitational radiation.

On the observational side gravitational radiation is known to exist thanks to
the discovery by Hulse and Taylor \cite{HT75} of the binary pulsar PSR 1913+16.
Five years of accurate timing observations of this pulsar have yielded the
conclusion (by Taylor, Fowler and Mc~Culloch \cite{TFMc79}) that the orbital
period of the binary system of the pulsar and its companion is steadily
decreasing, and therefore that by Kepler's third law the system is loosing
energy. There is exact agreement with the value of the flux of energy by
gravitational radiation as predicted by the Einstein quadrupole formula
\cite{PeM63}--\cite{Wag75}. This effect can also be seen as due to the reaction
forces acting on the orbit in reaction to the emission of gravitational
waves. Radiation reaction forces in general relativity and the associated
energy balance equation have been computed by Chandrasekhar and Esposito
\cite{CE70} (see also \cite{C65}--\cite{CN69}), Burke and Thorne
\cite{Bu69}--\cite{Bu71}, and many others authors \cite{AD75}--\cite{A87}.
None of these computations, however, gave a complete description of the
dynamics of gravitating systems, nor were applicable to systems
containing strongly self-gravitating bodies such as neutron stars.
The
first complete derivation of the dynamics of the binary pulsar, up to
the level where radiation reaction effects appear, was obtained by
Damour and Deruelle \cite{DD81a}--\cite{D83b}
(see the lecture of T.~Damour in this volume).

Experimental research on gravitational radiation is presently very active with
the development of new technologies for bar detectors, and most importantly
with the construction of two large scale laser interferometric detectors
for observation of the waves in the frequency bandwidth between $\approx$
10 Hz and $\approx$ 1000 Hz~: the American LIGO detector \cite{LIGO}, and
the Franco-Italian VIRGO detector \cite{VIRGO}. These detectors should
observe at least one type of source, the inspiralling compact binary.
This source is composed of two neutron stars or black holes
spiralling very rapidly around each other in the last rotations
preceding their final coalescence.  Orbital velocities are much larger
than in the binary pulsar for instance, which will coalesce with its
companion in few hundreds of millions years.

For very relativistic sources like inspiralling compact binaries, the precision
given by the Einstein quadrupole formula for the energy in the waves is
unsufficient. Similarly unsufficient is the precision of the radiation reaction
force of Burke and Thorne for instance. What is required is a relativistic or
{\it post-Newtonian} formalism (involving an expansion when the speed of light
goes to infinity) for both the emission and reaction of waves from isolated
sources with substantially large internal velocities. However as we do not
know what are the sources which will actually be detected by VIRGO and LIGO,
it is important to have a formalism which is sufficiently general
(not limited, for instance, to inspiralling compact binaries).

Pionneering work for extending the Einstein quadrupole formula (and formulas
alike) to post-Newtonian order is due to Epstein and Wagoner \cite{EW75}.
Note however that the post-Newtonian expansion is limited to the near-zone of
the source (see Section 2.1).
Thus, for studying the radiation field one needs a priori to supplement the
post-Newtonian expansion by some another method. The post-Minkowskian method
(expansion when the gravitational constant $G$ goes to zero) is adequate
for that purpose. Pionneering work on post-Minkowskian expansions is  due
to Bertotti and Plebanski \cite{BP60}, Bonnor and collaborators
\cite{Bo59}, \cite{BoR61}--\cite{Bo74}, and Thorne and collaborators
\cite{ThK75,CTh77}.  Notably Bonnor \cite{Bo59} introduced the idea of
considering the post-Minkowskian expansion outside the source where a
multipolar decomposition of the field simplifies the resolution of the field
equations.

Our aim in this lecture is to discuss, with emphasis on recent developments,
a particular post-Newtonian (and post-Minkowskian)
formalism for the generation of gravitational waves by a relativistic
source, and the reaction of the waves on the source.  Part of this
formalism is issued from the line of research initiated by Bonnor and
collaborators \cite{Bo59}, \cite{BoR61}--\cite{Bo74}.  Initial work was
inspired by Thorne's review on multipole expansions \cite{Th80}. The formalism
is sufficiently mature to tackle the problem of inspiralling compact binaries,
and we present at the end of the lecture the results which have been obtained
so far in this application.  The work on this post-Newtonian formalism
started with the present author's PhD thesis (advised by T.~Damour and
published in \cite{BD86}).  The list of references is provided at the end
of the lecture \cite{BD86}--\cite{B96tail}. A very useful auxiliary technical
development was brought by T.~Damour and B.~Iyer \cite{DI91b}.
Applications to binary systems (and, most
importantly, inspiralling compact binaries) were done in collaboration with
G.~Sch\"afer \cite{BS89,BS93}, T.~Damour and B.~Iyer \cite{BDI95}, and
partly with C.~Will and A.~Wiseman \cite{BDIWW95,BIWW95} who used also
their own method based on the Epstein and Wagoner approach \cite{EW75,WWi95}.
A. Wiseman \cite{Wi92,Wi93} and L.~Kidder {\it et al} \cite{KWW93,Kidder95}
made also applications of the formalism.

The present lecture is strongly biased by the
author's point of view, so we would like to refer here to other articles
for different approaches and methods, or simply for different points of views.
See in particular the Les Houches book of 1982 on Gravitational Radiation
\cite{Houches82} (chapters by Thorne, Damour, Walker, Choquet-Bruhat,
Friedrich, Ashtekar), the chapters of Thorne and Damour in the book on 300
years of gravitation \cite{300years} (see in particular the discussions on
approximation methods), the review by Damour in the Carg\`ese book on
Gravitation in Astrophysics \cite{Cargese} (see especially the discussion on
the various quadrupole laws and equations), the chapters by Iyer and Schmidt
in Advances in Gravitation and Cosmology \cite{ICGC91}, several reviews
by Thorne \cite{Th93}--\cite{Th94'}, Schutz \cite{Sc89',Sc93} and Bonazzola
and Marck \cite{BonaM94} more
devoted to detectors and the sources of waves, and one review by Will
\cite{W94} on the post-Newtonian approach to gravitational radiation.
See also the lectures of J.~Bi\v c\'ak and T.~Damour in this volume.

We will proceed didactically starting in Section II with several useful
definitions of regions around an isolated source, some recalls from the
quadrupole formalism, and a general outline of the method. Sections III and IV
are devoted respectively to some relevant properties of radiative gravitational
fields and the generation of waves by the source. Section V deals with
the radiation reaction effects occuring within the source. Finally Section VI
applies the results of this wave generation formalism to inspiralling compact
binaries.

\section{BASICS AND OUTLINE OF THE METHOD} \label{sec:2}

\subsection{Regions around an isolated source}\label{sec:2.1}
  
All over this lecture, we consider a general isolated source of gravity
described by a matter stress-energy tensor with components $T^{\mu\nu}
({\bf x},t)$ in some coordinate system $({\bf x},t)$ covering the
source (Greek indices take the values $0,1,2,3$, and Latin $1,2,3$).
Let $a$ be the radius of a sphere which totally encloses the source.

We shall make the restriction that this isolated source is slowly moving,
in the sense that the typical internal velocity inside the source divided by
the speed of light $c$, i.e. the typical ratio $T^{0i}/T^{00}$, defines a
small parameter $\varepsilon \approx 1/c$ referred to as the slow motion
or post-Newtonian parameter. We assume further that the typical stresses
inside the source divided by the energy density in the source, i.e. the
typical ratio $T^{ij}/T^{00}$, is of order of the square of the parameter
$\varepsilon$. Finally we consider a purely gravitational problem, so that the
source is self-gravitating (its internal motions are driven by gravitational
forces). In this case the ratio $G M/ c^2 a$, where $M$ is the total
mass-energy (or ADM mass) of the source, is also of order $\varepsilon^2$.
This implies that the gravitational field is weak (and
of order $\varepsilon^2$) everywhere inside and outside the source.

Note that the parameter $\varepsilon$ is required to be small, but not
exceedingly small. For instance it could be as large as say 0.2 or 0.3. This
is in order to allow for the very interesting case of inspiralling
compact binaries where $\varepsilon$ can reach 0.3 during the last rotations.
Thus we are here considering truly {\it relativistic} sources, whose accurate
description necessitates controlling many relativistic corrections
(how many depending on the exact range of magnitude of $\varepsilon$).
However, what we do not consider in this lecture is the case of a fully
relativistic source, which is made of massless particles
moving at the speed of light, or, to give another example, a cosmic string
whose internal stress is of the same order as its energy density.

A useful characterization of slowly moving sources is that their spatial
extension is small (and of order $\varepsilon$) as compared to one typical
wavelength of the radiation. Indeed the wavelength is given by
$\lambda = c P$ where $P$ is a period of motion in the source.
But $a\approx vP$ with $v\approx \varepsilon c$, thus $c P \approx a /
\varepsilon$ and therefore $a / \lambda \approx \varepsilon$ which is
indeed the statement above. Note that since $GM/ac^2\approx \varepsilon^2$
we have also $GM/\lambda c^2 \approx \varepsilon^3$.

It is convenient for slowly moving sources to introduce an interior domain
$D_i$, sometimes called also the near zone, defined by $D_i = \left\{
({\bf x},t),\, |{\bf x}| < r_i \right\}$, where the radius $r_i$ is such
that $r_i \approx \varepsilon\lambda$ {\it and} $r_i > a$.  The near zone
$D_i$ is small with respect to the wavelength of the radiation and
covers entirely the source.  This is possible only for a slowly moving
source. We choose $r_i$ to be strictly larger than $a$ (instead of
being $a$ itself) for later convenience.

The near-zone is the domain where one can confidently use the
post-Newton\-ian expansion.  The real precision of some post-Newtonian
expression will be exactly given, in $D_i$, by the formal order in
$\varepsilon$ of the neglected terms.  Note that all powers of $1/c$
must be taken into account in finding the magnitude of a term in terms of
$\varepsilon\approx 1/c$, including the ones which arise from the
temporal gradients $\partial_0 = c^{-1}\partial/\partial t$ (which are
really of order $\varepsilon$ with respect to the spatial gradients
$\partial_i$). This is clear from the definition of the near zone,
where the field is quasi-static and propagation effects are small.
These considerations are also familiar from electromagnetism.

Having clarified the concept of near-zone $D_i$, it is now necessary to
introduce an exterior domain defined by $D_e = \left\{ ({\bf x},t),|{\bf
x}| > r_e \right\}$, where the radius $r_e$ is chosen to be strictly between
$a$ and $r_i$, i.e. $a < r_e < r_i$. This choice, which can always be done
for slowly-moving sources, is in order that the intersection between
$D_i$ and $D_e$ (the exterior part of the near
zone $D_i \cap D_e$) exists.

Included in the exterior domain $D_e$, we also consider the so-called
exponentially far wave zone, namely the domain $D_w = \left\{
({\bf x},t), |{\bf x}| > r_\omega \right\}$, where $r_\omega$ is a radius
such that $r_\omega \approx \lambda e^{\lambda c^2/GM}$. In $D_w$, where
the observer will be located, one can expand the field in powers of the
distance of the source, and keep the leading order term in this
expansion.  The error done in assuming this will be negligible
for future astrophysical sources of radiation.

As is well-known from the study of the Schwarzschild solution, the actual
space-time light cones at large distances from the black hole deviate by
a logarithmic term from the ``flat'' light cones associated with some
coordinate systems more suitable near the source (for instance a harmonic
coordinate system, or a Schwarzschild-like one). The definition of $D_w$ as
being the exponentially far wave zone is chosen precisely in order that this
effect becomes appreciable there. Our exponentially far zone corresponds
to the distant wave zone in the terminology of Thorne \cite{Th80}, who
introduces also a local wave zone starting several wavelength apart from
the source. We shall not need here to make this distinction. It will be
sufficient to consider a minimal splitting into three regions $D_i$,
$D_e$ and $D_w$.  See Thorne \cite{Th80} for physical discussions of the
effects which arise in different regions, and for a more general
situation of a possibly strong-field source and a non-flat gravitational
background.

Let us comment that all the radius $a$, $r_i$, $r_e$ and $r_w$ which have
been introduced will in fact never appear explicitly in the results presented
below.  We shall only use the {\it existence} of the various domains as
determined by these radius, for instance the fact that the intersection
$D_i \cap D_e$ is not empty. This will permit us to prove some statements
about the construction of solutions, for instance that one can licitly match
the outer field to the inner field of the source in $D_i \cap D_e$, but
the precise locations of the radius are unimportant
(these radii are anyway loosely defined).

\subsection{The quadrupole formalism}  \label{eq:2.2}

By quadrupole formalism we mean the lowest order formalism, in the
post-Newtonian expansion $\varepsilon \to 0$, for the generation of
gravitational radiation from the source, and also for the reaction of the
radiation onto the source. Although this may seem to be a little paradoxical,
the quadrupole formalism can thus be viewed as a Newtonian formalism.
Relativistic corrections to the formalism are referred to as post-Newtonian
corrections accordingly to which post-Newtonian order in the equations
of motion of the source (beyond the usual Newtonian acceleration) is needed
in order to reduce all accelerations with consistent accuracy. For instance,
the first-post-Newtonian (in short 1PN) formalism retains all terms in the
radiation field and in the reaction which can be computed consistently using
the 1PN equations of motion, which include the corrections $\varepsilon^2$
beyond the Newtonian force (using the standard practice that the $n$th
post-Newtonian order refers to the order $\varepsilon^{2n}$).

The quadrupole formalism for the wave generation expresses the gravitational
field $h_{ij}^{\rm TT} = (g_{ij}-\delta_{ij})^{\rm TT}$ (where $g_{ij}$
denotes the spatial part of the covariant metric and $\delta_{ij}$ the
Kronecker symbol) in a transverse and traceless (TT) coordinate system as
\begin{equation}
 h^{\rm TT}_{ij} = {2G\over c^4R} {\cal P}_{ijkm} ({\bf N}) \left\{
 {d^2Q_{km}\over dT^2} (T-R/c) + O(\varepsilon)\right\}
   + O \left( {1\over R^2} \right)\ . \label{eq:1}
\end{equation}
The coordinate system $({\bf X},T)$ is centered on the source,
with $R = |{\bf X}|$ the distance to the source and ${\bf N} = {\bf X}/R$ the
unit direction from the source to the observer (later we shall need to be
more precise about the exact definition of this coordinate system). The
retarded time at the observer position in $D_w$ is $T-R/c$, and terms
of order $1/R^2$ in the distance of the source are neglected (in addition
to the post-Newtonian terms of order $\varepsilon$). In front of (1) appears
\begin{equation}
 {\cal P}_{ijkm} ({\bf N}) = (\delta_{ik} -N_i N_k) (\delta_{jm}-N_jN_m)
 -{1\over 2} (\delta_{ij} -N_iN_j) (\delta_{km}-N_kN_m) \label{eq:2}
\end{equation}
which is the TT projection operator onto the plane orthogonal to the direction
${\bf N}$. The quadrupole moment of the source takes the familiar Newtonian form
\begin{equation}
 Q_{ij} (t) = \int d^3 {\bf x}\ \rho ({\bf x},t) \left(x^i x^j -{1\over 3}
 \delta^{ij} {\bf x}^2 \right) \ , \label{eq:3}
\end{equation}
where $\rho$ denotes the Newtonian mass density of the source (in some
coordinate system which we shall also need to specify later). The
quadrupole moment (\ref{eq:3}) is taken to be tracefree ($\delta_{ij}
Q_{ij}=0$). This is not important in the waveform (\ref{eq:1}) because at the
quadrupolar level the TT operator in front cancels any
possible trace in the moment, but it will be very convenient in higher
orders to choose all multipole moments to be symmetric and tracefree
(STF) with respect to all their indices.

By differentiating with respect to time the waveform (\ref{eq:1})-(\ref{eq:3}),
squaring the result using standard formulas and integrating over all the
directions ${\bf N}$, one obtains the total power
in the gravitational waves emitted by the source,
\begin{equation}
 {\cal L} = {G\over 5c^5} \left\{ {d^3Q_{ij}\over dT^3} {d^3Q_{ij}\over dT^3} 
 + O(\varepsilon^2) \right\} \ . \label{eq:4}
\end{equation}
This result is the famous Einstein quadrupole formula \cite{E18}, which
neglects post-Newtonian terms of order
$\varepsilon^2$. It was derived originally by Einstein under the very
restrictive assumption that the motion of the source is non gravitational.
Then Landau and Lifschitz \cite{LL0,LL} showed that the quadrupole
formula applies also to a gravitationally bound system, for instance a
Newtonian binary star system. Many improvements in the derivation of this
formula have been obtained later, notably by Fock \cite{Fock} who made use of
the idea of matching between the inner field of the source and some exterior
field in $D_i \cap D_e$. See \cite{BD89} for a completely satisfactory
derivation of the quadrupole formula (and its first relativistic corrections)
using Fock's idea.  The notation ${\cal L}$ in (\ref{eq:4}) stands for
the total luminosity of the source in gravitational waves, by analogy
with the total luminosity of a star in electromagnetic waves.

It is straightforward using the expression of the quadrupole moment of the
source to rewrite the quadrupole formula (\ref{eq:4}) in the form of a balance
equation relating the mechanical energy loss in the source to the total
flux of energy in the waves. Indeed we can easily rewrite (\ref{eq:4})
in the form
\begin{equation}
 -{\cal L} + {G\over 5c^5} {d\over dT} \bigl[ Q_{ij}^{(3)} Q^{(2)}_{ij}
  - Q^{(4)}_{ij} Q^{(1)}_{ij} \bigr] = \int d^3 {\bf x} v^i
   {\cal F}^i_{\rm reac} + O(\varepsilon^7) \ . \label{eq:5}
\end{equation}
The superscript ($n$) denotes $n$ time derivatives. The second term in the
left-hand-side of (\ref{eq:5}), which has been added to the energy flux, is a
total time derivative and therefore its time average is numerically small in
the case of quasi-periodic motion (see e.g. \cite{BRu81} for a discussion).
On the other hand, we can interpret the right-hand-side as the power
extracted within the source by the radiation reaction force density
${\cal F}^i_{\rm reac}$, where ${\cal F}^i_{\rm reac}$ is given
explicitly by
\begin{equation}
 {\cal F}^i_{\rm reac} = \rho\ \partial_i V_{\rm reac} + O(\varepsilon^7)\ ;
\qquad V_{\rm reac} ({\bf x},t) = -{G\over 5c^5} \left\{ x^ix^j Q_{ij}^{(5)}(t)
      + O(\varepsilon^2) \right\}\ . \label{eq:6}
\end{equation}
This expression for the radiation reaction
force represents the gravitational analogue of the damping
force of electromagnetism, and was obtained by Burke and Thorne
\cite{Bu69}--\cite{Bu71}. Note that gravitational radiation reaction is
inherently gauge-dependent, so the expression of the force depends on the
coordinate system which is used (here, a Burke-Thorne coordinate system
whose definition can be found for instance in Chapter 36 of Misner, Thorne
and Wheeler \cite{MTW}). Chandrasekhar and Esposito \cite{CE70} have
independently obtained another (and more intricated) expression of the reaction
force, and Miller \cite{Mi74} has shown how one can transform one expression
into the other by means of a coordinate transformation (see Sch\"afer
\cite{S83} for a comparison between various expressions of the reaction
force).

When dealing with relativistic sources having $\varepsilon\, \alt\, 0.3$ (say),
the post-Newton\-ian terms which are neglected in the
quadrupole formulas both for the wave generation (\ref{eq:1})-(\ref{eq:4}) and
radiation reaction (\ref{eq:5})-(\ref{eq:6}) will become important.  Our
aim in this lecture is to compute these terms.

\subsection{Outline of the method}  \label{sec:2.3}

Generalizing the previous quadrupole formalism for application to relativistic
sources entails solving the mathematical problem of the resolution of the
Einstein field equations with due taking into account of all the
nonlinear effects
entering some high-order post-Newtonian approximation we are considering.

The general method we use can be decomposed into several different steps. As
a first step one investigates the gravitational field generated by the
isolated source in all the exterior domain $D_e$ (which comprises the far wave
zone $D_w$).  The field in $D_e$ is a solution of the vacuum field
equations and is computed using the combination of two methods~:

(i) a formal infinite post-Minkowskian approximation method for the metric
field (or perturbation around Minkowski's space-time, or also expansion
in powers of Newton's constant $G$) ---~this method is valid wherever the
field is weak~;

(ii) a decomposition of each coefficients of the latter post-Minkowskian
expansion into multipole moments (or simply expansion in spherical
harmonics), which is valid only in the exterior $D_e$.

The precise assumptions underlying the methods (i)-(ii) can be found in
Section 1.2 of \cite{BD86}. An iterative resolution ({\it \`a la} Bonnor
\cite{Bo59}) of the field equations permits us in principle to control, and
if necessary to compute, all the nonlinearities of the exterior field.
Spherical harmonics are expressed using the symmetric and tracefree
(STF) notation which is especially convenient when dealing with the
decomposition of tensors. This leads also to transparent definitions of the
multipole moments. See Thorne \cite{Th80} and the appendices A and B of
\cite{BD86} for many technical formulas and discussion.

The second step consists of

(iii) expanding the latter solution in the far wave zone $D_w$ (distance
$R\to\infty$) in order to find the observable moments of the radiative
field which are actually measured in the far away detector.

This step involves constructing a suitable (Bondi-type) coordinate system valid
in $D_w$ and correcting in particular for the logarithmic deviation of the true
light cones with respect to the flat light cones. The construction to all
orders in the post-Minkowskian expansion of such a coordinate system can be
found in \cite{B87}. This yields the operational link between the observable
moments and the moments parametrizing the multipole decomposition (ii).
Computations of the observable moments with increasing precision can be
found in the sections II.C of \cite{BD92}, IV.C of \cite{B95}, V of
\cite{B95pn}, and in \cite{B96tail}.

Finally it remains to bridge the field in $D_e$ to the inner field of the
source which is solution in $D_i$ of the non-vacuum Einstein field equations.
To do so we

(iv) compute the post-Newtonian expansion of the external post-Minkowskian
field and match it in the overlapping region $D_i \cap D_e$ to an inner
solution obtained by direct post-Newtonian iteration of the field
equations in $D_i$.

This is an application of the method of matched asymptotic
expansions \cite{Lagerst}. See \cite{BuTh70} for a discussion on how useful
is this method for bridging the near zone and the exterior zone. After
matching, the field is determined everywhere inside and outside the source
as a functional of the source's parameters. This allows the determination
both of the exterior multipole moments as integrals over the source, and
of the radiation reaction forces acting inside the source. Matchings have
been performed in the sections III of \cite{BD89}, IV of \cite{DI91a} and III
of \cite{B95} to find the multipole moments, and in the sections VI of
\cite{BD88} and in \cite{B95reac} to find the reaction forces.

\section{PROPERTIES OF RADIATIVE GRAVITATIONAL FIELDS} \label{sec:3}

\subsection{The field in the exterior domain $D_e$} \label{sec:3.1}

In $D_e$ we have the vacuum field equations $R_{\mu\nu}[g]=0$. It is
advantageous to use
instead of the covariant
metric $g_{\mu\nu}$ (with determinant $g$) the densitized metric
deviation $h^{\mu\nu} =\sqrt{-g} g^{\mu\nu} -\eta^{\mu\nu}$
as a basic field variable,
where
$\eta^{\mu\nu}={\rm diag}(-1,1,1,1)$ is the Minkowski metric. The
convention is that the indices on $h^{\mu\nu}$ (but not on $g^{\mu\nu}$
which is the inverse of $g_{\mu\nu}$) are lowered using the flat metric
$\eta_{\mu\nu} = {\rm diag} (-1,1,1,1)$. With this convention one defines
for instance $h_{\mu\nu}=\eta_{\mu\rho} \eta_{\nu\sigma} h^{\rho\sigma}$
and $h=\eta_{\mu\nu}h^{\mu\nu}$. Imposing the harmonic (or De Donder)
gauge condition for the coordinates $x^\mu$ (considered as forming four
scalars $x^0,\cdots, x^3$) yields $0 = \Box_g x^\mu = - g^{\rho\sigma}
\Gamma^\mu_{\rho\sigma} = - (-g)^{-1/2} \partial_\nu h^{\mu\nu}$, hence
\begin{equation}
 \partial_\nu h^{\mu\nu} = 0 \ . \label{eq:7}
\end{equation}
The vacuum field equations in these coordinates then read
\begin{equation}
  \Box h^{\mu\nu} = \wedge^{\mu\nu} (h)\ ,  \label{eq:8}
\end{equation}
where $\Box \equiv \Box_\eta = \eta^{\mu\nu} \partial_\nu \partial_\nu$ denotes
the Minkowski (flat) d'Alembertian operator, and where $\Lambda^{\mu\nu}
(h)$ can be viewed as a source for the gravitational field in vacuum which
originates from the gravitational field itself.  We have $\partial_\nu
\Lambda^{\mu\nu} =0$ as a consequence of the Bianchi identities, which
is implied here by the harmonic coordinates condition.

The source $\Lambda^{\mu\nu} (h)$ encompasses all the
nonlinearities of the field equations, and is at least quadratic in $h$ and
its first and second space-time derivatives. For instance the quadratic
part of $\Lambda^{\mu\nu} (h)$, which is such that $\Lambda^{\mu\nu}
(h)=N^{\mu\nu} (h,h) + O(h^3)$, is explicitly given by
$(t^{(\mu\nu)}={1\over 2} (t^{\mu\nu}+t^{\nu\mu}))$
\begin{eqnarray}
 N^{\mu\nu} (h,h)&=&- h^{\rho\sigma} \partial_\rho \partial_\sigma
 h^{\mu\nu} + {1\over 2} \partial^\mu h_{\rho\sigma} \partial^\nu
 h^{\rho\sigma} - {1\over 4} \partial^\mu h \partial^\nu h \nonumber\\
&&-2 \partial^{(\mu} h_{\rho\sigma} \partial^\rho h^{\nu)\sigma}
  +\partial_\sigma h^{\mu\rho} (\partial^\sigma h^\nu_\rho + \partial_\rho
  h^{\nu\sigma}) \nonumber \\
&&+ \eta^{\mu\nu} \left[ -{1\over 4}\partial_\lambda h_{\rho\sigma}
  \partial^\lambda h^{\rho\sigma} +{1\over 8}\partial_\rho h \partial^\rho h
 +{1\over 2}\partial_\rho h_{\sigma\lambda} \partial^\sigma h^{\rho\lambda}
   \right]\ .  \label{eq:9}
\end{eqnarray}
The cubic part of $\Lambda^{\mu\nu} (h)$ can be found e.g. in the section
I.B of \cite{B95}.

The main technique we use in this section is to look for the solution of
the field equations in the form of the post-Minkowskian expansion
\begin{equation}
  h^{\mu\nu} = G \, h^{\mu\nu}_1 + G^2 \, h^{\mu\nu}_2 + \cdots + G^n \,
    h^{\mu\nu}_n + \cdots \ ,\label{eq:10}
\end{equation}
where we denote by $G$ an ordering (or book-keeping) parameter which shall be
identified later with Newton's constant.  The mathematical status of the
post-Minkowskian expansion is known in vacuum (as is the case here). This
expansion is asymptotic, in the sense that any solution of the perturbation
equations (written in (\ref{eq:11})-(\ref{eq:12}) below)
comes from the Taylor expansion when $G\to 0$ of a family of exact
solutions \cite{DS90}.  Therefore we do not loose any generality in
assuming the expansion (\ref{eq:10}), provided that this expansion
is carried out up to infinity.

Substituting the asymptotic expansion (\ref{eq:10}) into the vacuum Einstein
equations (\ref{eq:7})-(\ref{eq:8}), and equalizing all the coefficients of
the $G^n$'s in both sides of the equations, one obtains an infinite ($n = 1,
\cdots, \infty$) set of perturbation equations to be solved for all the
$h^{\mu\nu}_n$'s. These equations take the form of wave equations in the
ordinary sense supplemented by the condition of harmonic coordinates,
\begin{equation}
 \Box h^{\mu\nu}_n = \Lambda^{\mu\nu}_n (h_1 , \cdots , h_{n-1})\ ;
  \quad \partial_\nu h^{\mu\nu}_n = 0 \ , \label{eq:11}
\end{equation}
where the important point is that since $\Lambda^{\mu\nu}$ is at least
quadratic in $h$ (see (\ref{eq:9})), all the nonlinear sources
$\Lambda^{\mu\nu}_n$ depend for any given $n$ only on the previous
iterations $h^{\mu\nu}_m$ with $1 \leq m \leq n-1$.  For instance we
have $\Lambda^{\mu\nu}_2=N^{\mu\nu}(h_1,h_1)$ where $N^{\mu\nu}$ is
given by (\ref{eq:9}).

When $n=1$, the corresponding source is zero, $\Lambda^{\mu\nu}_1 = 0$, and
therefore the coefficient of $G$ in the expansion (\ref{eq:10}), which
corresponds to the linearized approximation, satisfies
\begin{equation}
\Box h^{\mu\nu}_1=0\quad ;\quad\partial_\nu h^{\mu\nu}_1 = 0\ . \label{eq:12}
\end{equation}
An elementary retarded solution in $D_e$ of the source-free wave equation
$\Box \varphi=0$ reads $\varphi =F(t-r/c)/r$, where $F$ is an arbitrary
function of the retarded time $t-r/c$
(with $r=|{\bf x}|$ the distance to the spatial origin of coordinates located 
in the source). The most general retarded solution is readily obtained by 
differentiating with respect to space the elementary solution an arbitrary
number of times, say $\ell$~: $\varphi_{i_1\cdots i_\ell} =
\partial_{i_1} \cdots \partial_{i_\ell}
\{ F(t-r/c)/r\}.$ We shall extensively use below the notation
$\partial_L \equiv \partial_{i_1} \cdots \partial_{i_\ell}$ to denote a
product of $\ell$ space derivatives, where $L=i_1 i_2\cdots i_\ell$ represents
a multi-index composed of $\ell$ indices. With this notation $\varphi_L=
\partial_L \{F(t-r/c)/r\}$.  Thorne \cite{Th80} has shown that the most
general solution $h^{\mu\nu}_1$ of the equations (\ref{eq:12}) can always
be written, modulo an infinitesimal gauge transformation preserving the
harmonic gauge condition, in terms of two and only two types of arbitrary
functions $F_L$, which can be chosen to be symmetric and tracefree (STF)
with respect to the $\ell$ indices they carry (i.e. $F_L$ is symmetric
under the permutation of any pair of its indices, and $\delta_{i_1i_2}
F_L =0$).  We shall denote by $M_L(t)$ and $S_L(t)$ these two types of
STF tensorial functions of time and we shall call them the ``algorithmic''
multipole moments of mass type $(M_L,$ having $\ell =0,1, \cdots)$ and of
current type $(S_L,$ having $\ell =1,2,\cdots)$. (The adjective ``algorithmic''
refers to the algorithmic construction below of the external field.)  In the
time-independent case (for stationary systems), these multipole moments
are identical, aside from normalization, to the Geroch \cite{G70} and
Hansen \cite{H74} multipole moments (see the proof in \cite{Gu83}).

The moments $M_L(t)$ and $S_L(t)$ are for the moment arbitrary except that if
the harmonic gauge condition in (\ref{eq:12}) is to be satisfied, the lowest
order moments $M$ and $M_i$ (mass monopole and dipole moments), and $S_i$
(current dipole moment) are necessarily constant. This corresponds to the laws of
conservation of the total mass, of the position of the center of mass, and of
the total angular momentum (the linear momentum which is the time derivative of
$M_i$ is zero because we consider the case of a source which was at rest in the
infinite past). Furthermore we shall often choose $M_i=0$ by translating the
origin of the coordinates to the center of mass. The mass monopole $M$ is the
total ADM mass \cite{ADM} of the source.

We shall write the result for $h^{\mu\nu}_1$ in the form
\begin{eqnarray}
 G\,h^{00}_{1}= - {4\over c^2}\,V^{\rm ext}\ ; \quad
 G\,h^{0i}_{1}= - {4\over c^3}\,V_i^{\rm ext}\ ; \quad
 G\,h^{ij}_{1}= - {4\over c^4}\,V_{ij}^{\rm ext}\ ,\label{eq:13}
\end{eqnarray}
where we have introduced for later convenience some external
potentials (valid in $D_e$) defined by
\begin{eqnarray}
 V^{\rm ext} &=& G \sum^\infty_{\ell =0} {(-)^\ell\over \ell !}
\partial_L \left[ {1\over r} \,M_L \left( t-{r\over c}\right)\right]\ ,
   \nonumber \\ 
 V_i^{\rm ext} &=& -G \sum^\infty_{\ell =1} {(-)^\ell\over \ell !}
  \partial_{L-1} \left[ {1\over r} \,M^{(1)}_{iL-1}
  \left( t-{r\over c}\right)\right]\nonumber \\
   && -G \sum^\infty_{\ell =1} {(-)^\ell\over \ell !}{\ell\over \ell +1}
 \varepsilon_{iab} \partial_{aL-1} \left[ {1\over r} \,S_{bL-1}
  \left( t-{r\over c}\right)\right]\ , \nonumber \\ 
 V_{ij}^{\rm ext} &=& G \sum^\infty_{\ell =2} {(-)^\ell\over \ell !}
  \partial_{L-2} \left[ {1\over r} \,M^{(2)}_{ijL-2}
  \left( t-{r\over c}\right)\right]\nonumber \\
   && +G \sum^\infty_{\ell =2} {(-)^\ell\over \ell !}{2\ell\over \ell +1}
\partial_{aL-2} \left[ {1\over r} \varepsilon_{ab(i}  \,S^{(1)}_{j)bL-2}
  \left( t-{r\over c}\right)\right]\ . \label{eq:14}
\end{eqnarray}
The $\ell$-dependent coefficients in these expressions are chosen
so that $M_L(t)$ and $S_L(t)$ reduce in the limit $\varepsilon
\to 0$ corresponding to a Newtonian source to the usual expressions of
the Newtonian moments of the source \cite{Th80}.  For instance,
\begin{equation}
  M_{ij} (t) = Q_{ij} (t) + O(\varepsilon^2)\ ,  \label{eq:15}
\end{equation}
where $Q_{ij}$ is the Newtonian mass quadrupole (\ref{eq:3}). A complete and
self-contained derivation of the result (\ref{eq:13})-(\ref{eq:14}) for
the linearized metric can be found in Section 2 of \cite{BD86}.

\subsection{Computing the field nonlinearities in $D_e$}\label{sec:3.2}

Suppose that starting from the linearized metric (\ref{eq:13})-(\ref{eq:14})
one has succeeded in solving the field equations (\ref{eq:11}) for all the
nonlinear coefficients $h^{\mu\nu}_m$ up to some given order $n-1$
$(1\leq m\leq n-1)$. Each of these coefficients $h^{\mu\nu}_m$ will be in
the form of a multipole expansion like $h^{\mu\nu}_1$.  How can we find
the solution of the equations for the next-order coefficient $h^{\mu\nu}_n$~?
The solution naturally involves the inverse retarded
d'Alembertian operator $\Box^{-1}_R$ defined by
\begin{equation}
 (\Box^{-1}_R f)({\bf x},t) = -{1\over 4\pi} \int\!\!\!\int\!\!\!\int
 {d^3{\bf x}'\over
 |{\bf x}-{\bf x}'|} f ({\bf x}', t-{1\over c} |{\bf x}-{\bf x}'|)\ ,
\label{eq:16}
\end{equation}
which satisfies $\Box(\Box^{-1}_R f) = f$. However in trying to solve the
wave equation in (\ref{eq:11}) directly with this operator one encounters a
problem. Namely the source term $\Lambda^{\mu\nu}_n$ is like the previous
coefficients $h^{\mu\nu}_m$ in the form of a multipole expansion which is
valid only in $D_e$, while the retarded integration in (\ref{eq:16}) extends
over all space, including the inner domain $D_i$. At the spatial origin of the
coordinates $r \to 0$ located in $D_i$, the multipole expansion is singular
(see the equations (\ref{eq:14})). This problem is of course not a physical
problem but simply a mathematical one, as we are simply interested in finding
a solution of the equations (\ref{eq:11}) which is legitimate in $D_e$.

As a cure of this problem we use a mathematical trick which allows us to find
the solution $h^{\mu\nu}_n$ of the equations (\ref{eq:11}) in the form of a
multipole expansion, while keeping the range of integration in the retarded
integral as it is in (\ref{eq:16}). This trick consists of replacing the real
source $\Lambda^{\mu\nu}_n$ by a fictituous source obtained by multiplying
$\Lambda^{\mu\nu}_n$ by $r^B$ where $B$ is a complex number. Assuming that the
real part of $B$ is large enough (and assuming that the multipole expansion is
actually finite), one ``removes'' in this way the singularity at the origin
$r \to 0$. There is no problem with the behaviour when $r\to +\infty$ if one
assumes that the source was stationary in the remote past. Then one can apply
to this fictituous source the operator (\ref{eq:16}) as it stands. The result
is in general infinite at $B=0$, but the point is that the retarded integral
of the fictituous source has been proved in Section 3 of \cite{BD86} to admit
(after analytic continuation) a Laurent expansion near $B \to 0$ whose finite
part, defined to be the coefficient of the zeroth power of $B$ in the expansion,
is precisely the solution of the wave equation in (\ref{eq:11}) we are looking
for. Denoting the finite part at $B=0$ by the symbol FP$_{B=0}$ we pose
\begin{equation}
 p^{\mu\nu}_n={\rm FP}_{B=0}\Box^{-1}_R(r^B\Lambda^{\mu\nu}_n)\ . \label{eq:17}
\end{equation}
Then $p^{\mu\nu}_n$ satisfies $\Box p^{\mu\nu}_n = \Lambda^{\mu\nu}_n$ and
is in the form of a multipole expansion like the previous coefficients
$h^{\mu\nu}_m$.  However the second equation in (\ref{eq:11}) which is
the harmonic gauge condition is still to be satisfied.  We shall refer
here to \cite{BD86} for the computation from the object $p^{\mu\nu}_n$
(or rather from its non-zero divergence $w^{\mu}_n \equiv \partial_\nu
p^{\mu\nu}_n$) of a second object $q^{\mu\nu}_n$ satisfying at once
$\Box q^{\mu\nu}_n = 0$ and $\partial_\nu q^{\mu\nu}_n = - w^{\mu}_n$
(see Eqs.(4.13) in \cite{BD86} for the defining expression of
$q^{\mu\nu}_n$).  With these definitions a solution of both the wave
equation and the gauge condition in (\ref{eq:11}) reads
\begin{equation}
  h^{\mu\nu}_n = p^{\mu\nu}_n + q^{\mu\nu}_n \ . \label{eq:18}
\end{equation}
This completes the construction of the $n$th coefficient of the
post-Minkowskian expansion (\ref{eq:10}), and therefore by induction of
a whole infinite asymptotic expansion solving the vacuum field
equations in $D_e$.

A priori, the above construction seems to be particular, however it has been
proved in Section~4.2 of \cite{BD86} to represent in fact the most
general solution of the field equations in $D_e$ (modulo an arbitrary change of
coordinates in $D_e$). The conclusion is that the general radiative field
outside an isolated system depends on two and only two sets of time-varying
multipole moments $M_L(t)$ and $S_L(t)$~: $h^{\mu\nu} = h^{\mu\nu}[M_L, S_L]$.
As a side result but whose proof is in fact crucial in the above construction
\cite{BD86}, it is shown that the general structure of the post-Newtonian
expansion $(\varepsilon \to 0)$ of the exterior field in $D_i \cap D_e$
is of the form $\varepsilon^{k}(\ln \varepsilon)^p$, where $k$ et $p$ are
positive integers. This result can be extended by a matching argument to the
whole near zone $D_i$. Using some energy and momenta balance arguments  the
result could also (probably) be extended to the post-Newtonian expansion
of the radiative field in $D_w$.  The exact level at which the first
logarithms of $\varepsilon$ appear is investigated in \cite{BD88} (see also
\cite{A79}--\cite{Fu83}).

In principle the construction of the field in $D_e$ could be implemented by an
algebraic computer programme. However more theoretical work should be done
first on the occurence and properties of high-order tail integrals in
high-order post-Minkowskian approximations (a preliminary investigation
at the cubic level can be found in \cite{B96tail}).
 
For practical computations one needs some explicit formulas expressing the
retarded integral (\ref{eq:16}) when the function $f$ has a definite
multipolarity $\ell$, say $f({\bf x},t)= \hat n_L f(r,t)$ where $\hat
n_L$ denotes the STF part of the product of unit vectors $n_L=n_{i_1}\cdots
n_{i_\ell}$ ($n_i=x^i/r$). As reviewed in \cite{Th80} and Appendix A of
\cite{BD86}, the STF tensors $\hat n_L=\hat n_L(\theta,\phi)$ can be used
equivalently to the usual spherical harmonics $Y_{\ell m}(\theta,\phi)$ for
the multipole decomposition.  The relevant formulas for this are
(6.3)--(6.4) in \cite{BD86}, namely
\begin{equation}
 \Box^{-1}_R (\hat n_L f(r,t)) = \int^{t-r/c}_{-\infty} cds \,
  \hat\partial_L \left\{ {g\left( {t-r/c-s\over 2},s \right)
   - g \left( {t+r/c-s\over 2},s \right) \over r} \right\}\ , \label{eq:19}
\end{equation}
where the function $g$ is related to $f$ by
\begin{equation}
  g(\rho, s) = \rho^\ell \int^{\rho}_\alpha dx\, {(\rho -x)^\ell\over \ell !}
    \left( {2\over x}\right)^{\ell-1} f(x,s+x/c) \  \label{eq:20}
\end{equation}
(where $\alpha$ is arbitrary).
The formulas (\ref{eq:19})-(\ref{eq:20}) are correct only when the source
$f(r,t)$ tends to zero when $r \to 0$ more rapidly than some power of $r$
(specified in \cite{BD86}). This will always be the case of the fictituous
source ${}_B f = r^B f$ which was used in the previous construction of
the exterior field (taking $Re(B)$ initially large enough and analytically
continue the result). Notice the nice correspondence in (\ref{eq:19})
between the STF tensor $\hat n_L$ in the source of the wave equation and
the STF derivative operator $\hat \partial_L$ (STF part of the product
of derivatives $\partial_{i_1} \cdots \partial_{i_\ell}$) in the
corresponding solution.

Here are some formulas derived from (\ref{eq:19})-(\ref{eq:20}) in Section 4
of \cite{BD88} and which are particularly useful in the study of the
quadratic nonlinearities, in which case one needs to
integrate some source terms of the form $f(r,t)=r^{-k}F(t-r/c)$ with $k \geq 2$.
When $k=2$ we find
\begin{eqnarray}
&&\Box^{-1}_R \left[ \hat n_L r^{-2} F \left( t-{r\over c}\right) \right]
={(-)^\ell\over 2\ell !}\int^{+\infty}_r d\lambda\, F(t-\lambda/c)
\qquad\qquad \qquad\nonumber\\
 &&\qquad\qquad\qquad\quad \times \hat\partial_L \left\{ {(\lambda-r)^\ell
   \ln (\lambda-r) -(\lambda+r)^\ell \ln (\lambda+r) \over r}\right\}\ ,
  \qquad\label{eq:21}
\end{eqnarray}
which involves a non-local integral extending over the whole past of the
source (see below for the associated physical effects). On the
contrary the case $3 \leq k \leq \ell+2$ yields a local result~:
\begin{eqnarray}
&&\Box^{-1}_R \left[ \hat n_L r^{B-k} F \left( t-{r\over c}\right) \right]
_{|_{B=0}} = -{2^{k-3} (k-3)! (\ell+2-k)!\over (\ell+k-2)!}\qquad\qquad\nonumber \\
 &&\qquad\qquad\qquad\qquad\qquad\quad \times\hat n_L \sum^{k-3}_{i=0}
{(\ell+i)!\over 2^ii!(\ell-i)!}{F^{(k-3-i)}(t-r/c)\over c^{k-3-i}r^{1+i}}\ .
 \qquad\qquad\label{eq:22}
\end{eqnarray}
In these two cases the result is in fact finite at $B=0$ and therefore we have
dropped the Finite Part symbol.  When $k \geq \ell +3$ the result is no longer
finite (there is a simple pole at $B=0$) and the expression of the
finite part is more complicated. See \cite{B96tail} for other formulas enabling
one to investigate the cubic nonlinearities of the field equations.

\subsection{The field in the far wave zone $D_w$} \label{sec:3.3}

The study of ``asymptotics'' (namely what is the structure of space-time at
large distances from the source ?) is an active research field in general
relativity since the work of Bondi {\it et al} \cite{BBM62} and Sachs
\cite{S62} showing that it is possible to find some solutions of the field
equations with bounded sources in the form of a power series in the inverse of
the distance $R$ to the source (with a null coordinate $u$ remaining constant).
The reformulation of this approach in geometrical terms by Penrose
\cite{P63,P65} led to the concept of an asymptotically simple spacetime, which
is by definition a spacetime sharing with Minkowski's spacetime some
(mathematically precise) global and local asymptotic properties, and which we
would like to be associated with any radiating isolated system (see
\cite{G77}--\cite{A84} for reviews, and the lecture of J.~Bi\v c\'ak in
this volume).

The harmonic coordinate system used for the previous construction of the field
in $D_e$ was very convenient (because the equations for all components
of the field take the form of wave equations), but it has one drawback~:
it is not of the ``Bondi-type'' because the expansion of the field when
$r\to +\infty$ with $t-r/c =\ {\rm const}$ involves besides the normal powers
of $r^{-1}$ some logarithms of $r$.  This was noticed long ago by Fock
\cite{Fock}.  Madore \cite{Mad} proved that the harmonic-coordinates
linearized metric (\ref{eq:13})-(\ref{eq:14}) cannot be the first
approximation of a Bondi-type expansion at infinity.  The origin of the
problem is in the deviation of the flat light cones in harmonic coordinates
with respect to the true null cones.  The logarithms of the distance
present in $D_w$ imply that the $n$th post-Minkowskian approximation
there becomes larger than the $(n-1)$th one.  However this is not a real
problem because the logarithms of $r$ are not genuine -- they can be
removed by a change of coordinates in $D_w$ (in contrast with the
logarithms of $c$ in the post-Newtonian expansion which are probably
present in all coordinate systems covering $D_i$).

The construction of Bondi-type (sometimes also called radiative) coordinates
$X^\mu$ from the
harmonic coordinates $x^\mu$ has been achieved in the present framework to
all orders in the post-Minkowskian expansion \cite{B87}. A ``radiative''
post-Minkowskian metric was constructed first in the manner of
the previous section, but correcting at each post-Minkowskian
order the flat cones associated with the coordinate system in order that
they agree (at least asymptotically) with the real light cones $u={\rm const}$.
Then it was shown (section 4 of \cite{B87}) that the radiative metric so
constructed indeed differs from the general harmonic-coordinates metric by a
mere coordinate transformation.  As a corollary it was proved that the
Penrose definition of asymptotic simplicity (in a form given by Geroch and
Horowitz \cite{GH78}) is formally satisfied for any isolated radiating
source. Note that this proof holds only in the case
where the source is stationary in the past, i.e. when there exists a finite
initial instant $-{\cal T}$ at which the source starts vibrating and emitting
the radiation. If this assumption is relaxed, the definition is probably no
longer valid \cite{BaPr73}--\cite{D86'} (at least in its original form).

The main interest of the construction of Bondi-type coordinates in \cite{B87}
is that it is explicit and can be performed to all orders in $G$.
To leading order in $G$ one finds that the retarded time $T-R/c$ in Bondi-type
coordinates ($T-R/c$ and the null coordinate $u$ coincide in the limit $R \to
\infty$) differs from the harmonic coordinate time $t-r/c$ by
\begin{equation}
  T - {R\over c} = t - {r\over c} - {2GM\over c^3} \ln
     \left({r\over cb}\right) + O(G^2)\ . \label{eq:23}
\end{equation}
This is simply the known logarithmic deviation of light cones in harmonic
coordinates. The constant $b$ is an arbitrary gauge-dependent time
scale. The correction of order $G^2$ will be given below.

The Bondi-type coordinates are well-adapted to the study of the field in $D_w$, 
and most importantly permit a clear definition of two sets
of multipole moments $U_L(t)$ and $V_L(t)$ parametrizing the radiative field
at infinity, and which are directly ``measured'' in a detector located in
$D_w$. We shall refer to these moments as the ``observable'' or ``radiative''
moments, respectively of mass-type ($U_L$) and of current-type ($V_L$).
These moments are distinct from the algorithmic moments $M_L$ and
$S_L$. Their definition goes as follows. For all practical purposes the
distance of the source is so large (say $R$~=100~Mpc for inspiralling
binaries) that one can consider only the leading-order term $1/R$ in the
Bondi expansion. Like in (1) the wave is entirely characterized by the
TT projection of the spatial metric. It is a standard result (see e.g.
\cite{Th80}) that this projection can be decomposed uniquely into
multipole moments according to
\begin{eqnarray}
&&h^{\rm TT}_{ij} ({\bf X},T) = {4G\over c^2R} {\cal P}_{ijkm} ({\bf N})
 \sum^\infty_{\ell =2} {1\over c^\ell \ell !} \biggl\{ N_{L-2} U_{kmL-2}\!
 \left( T-R/ c\right) \nonumber \\
 && \qquad \qquad\quad- {2\ell \over (\ell +1)c} N_{aL-2}
 \varepsilon_{ab(k} V_{m)bL-2}\!\left( T-R/ c\right)\biggr\} +
 O \left( {1\over R^2}\right) .\ \label{eq:24}
\end{eqnarray}
The TT operator ${\cal P}_{ijkm}$ is given by (\ref{eq:2}), and $t_{(km)}
={1\over 2} (t_{km}+t_{mk})$. The multipole
decomposition (\ref{eq:24})
is the defining expression for the observable moments $U_L$ and $V_L$
(which are functions of the retarded time $T-R/c$). The total power, or
luminosity, contained in the waveform (\ref{eq:24}) is given by
\begin{eqnarray}
 {\cal L} &=& \sum^{+\infty}_{\ell=2} {G\over c^{2\ell+1}} \biggl\{
   {(\ell+1)(\ell+2)\over (\ell-1)\ell \ell!(2\ell+1)!!} U^{(1)}_L
   U^{(1)}_L \qquad\qquad \nonumber \\
 &&\qquad\qquad\qquad
  +{4\ell (\ell+2)\over (\ell-1)(\ell+1)!(2\ell+1)!!c^2}
   V^{(1)}_L V^{(1)}_L \biggr\} \ . \label{eq:25}
\end{eqnarray}
The $\ell$-dependent coefficients in (\ref{eq:24})-(\ref{eq:25}) are ajusted
so that the observable moments $U_L$ and $V_L$ agree at the linearized order
with the $\ell$th time derivatives of the algorithmic moments $M_L$ and $S_L$
(beware of the fact that $U_L$ and $V_L$ do not have the usual dimensions
of multipole moments). Comparing (\ref{eq:24}) with (\ref{eq:13})-(\ref{eq:14})
(and taking into account a sign difference due to the use in
(\ref{eq:13})-(\ref{eq:14}) of the metric perturbation $\sqrt{-g} g^{ij}
-\delta^{ij}$ instead of $g_{ij} -\delta_{ij}$ in (\ref{eq:24})) we
have \cite{Th80}
\begin{eqnarray}
 U_L (T) &=& d^\ell M_L /dT^\ell + O(G)\ , \nonumber \\  
 V_L (T) &=& d^\ell S_L /dT^\ell + O(G)\ , \label{eq:26}
\end{eqnarray}
where the terms $O(G)$ symbolize the nonlinear corrections
(which can be computed in principle to all orders using \cite{B87}). On
the other hand we have seen in (\ref{eq:15}) that the mass quadrupole moment
$M_{ij}$ agrees in the Newtonian limit with the quadrupole $Q_{ij}$ given
by the integral over the source (\ref{eq:3}). Thus, from (\ref{eq:26})
and (\ref{eq:15}), we see that in the Newtonian limit the connection
between the observable moment at infinity $U_{ij}$ and the actual source
moment $Q_{ij}$ is realized (nonlinear corrections in (\ref{eq:26})
vanish in the Newtonian limit). Of course this yields simply the
quadrupole formalism reviewed in Section~2.

What we shall do in the next subsection is to generalize the relationships
linking the observable multipole moments to the algorithmic moments in order to
include explicitly the nonlinear corrections of order $G$ (at least) appearing
in (\ref{eq:26}). The computation of the algorithmic moments themselves as
explicit integrals over the source generalizing (\ref{eq:15}) to
post-Newtonian order will be dealt with in the section~4.  It is clear
that knowing both the relations between the observable moments
$U_L$,$V_L$ and the moments $M_L$,$S_L$, and between the moments
$M_L$,$S_L$ themselves and the
source, one solves the problem at hand.

\subsection{Nonlinear effects in the radiation field} \label{sec:3.4}

It follows from their construction \cite{B87} that the observable moments
$U_L$ and  $V_L$ are necessarily given by some nonlinear functionals of the
moments $M_L$ and $S_L$ (reducing to (\ref{eq:26}) at the linearized
order).  A very useful information can be inferred on dimensional grounds.
Indeed the general {\it form} of the relations (\ref{eq:26}) can be very
easily proved to be
\begin{eqnarray}
 U_L(T) &=&M^{(\ell)}_L (T) + \sum_{n\geq 2} {G^{n-1}\over c^{3(n-1)+2k}}
  X_{nL} (T)\ , \nonumber \\ 
 \varepsilon_{ai_\ell i_{\ell-1}} V_{aL-2}(T)
  &=&\varepsilon_{ai_\ell i_{\ell-1}}
  S^{(\ell-1)}_{aL-2} (T) + \sum_{n\geq 2} {G^{n-1}\over c^{3(n-1)+2k}}
  Y_{nL} (T) \label{eq:27}
\end{eqnarray}
(endowing the current moment with its natural Levi-Civita symbol), where the 
functions $X_{nL}$ and $Y_{nL}$ represent some nonlinear functionals
of $n$ moments $M_L$ and $S_L$ whose general structure reads
\begin{eqnarray}
&& X_{nL} (T),\ Y_{nL} (T)=\qquad\qquad\qquad\qquad\nonumber \\
&&\sum \int^T_{-\infty} dv_1 \cdots \int^T_{-\infty}
 dv_n {\cal X} (T, v_1,\cdots, v_n) M_{L_1}^{(a_1)} (v_1)
  \cdots S^{(a_n)}_{L_n} (v_n)\ . \label{eq:28}
\end{eqnarray}
The (a priori) quite complicated kernel ${\cal X}$ has an index structure
made of Kronecker and Levi-Civita symbols, and depends only on variables
having the dimension of time. Accordingly the powers of $c^{-1}$ in each
$n$th nonlinear terms of (\ref{eq:27}) are such that the dimensions of the
moments $U_L$ and $V_L$ are respected. They turn out (see e.g. the
section~IV.C of \cite{B95}) to be of the type $3(n-1)+2k$ where $k$ is a
positive integer representing a number of contractions between the
indices on the moments composing the term in question.
This information is useful when determining the type of nonlinear
terms which arise in a formula like (\ref{eq:38}) below.

As expressed in (\ref{eq:28}) the functionals $X_{nL}$ and $Y_{nL}$ are
given by some highly non-local in time (or ``hereditary'') functionals of
the moments $M_L$ and $S_L$. This illustrates the fact that the gravitational
field in higher approximations depends on all the integrated past (or
``history'') of the source \cite{BP60}, \cite{YCB1}--\cite{YCB90}. The
propagation of the radiation proceeds not only along the characteristic
surfaces of spacetime or light cones, but also inside them. One can view
this phenomenon as the propagation at the speed of light of a front wave
spreading by continuous scattering a secondary wave propagating in average
at a speed less than $c$. This wave is referred generally to as the tail
of the wave. Perhaps a better name would be the wake of the wave. An
interesting particular case is that of the scattering onto the static
spacetime curvature generated by the total mass-energy $M$ of the source
itself. Often the name of tail of the wave is reserved for this specific
effect. In that case the tail results from the nonlinear interaction between
the varying multipole moments describing the radiation field, primarily
the quadrupole moment, and the static mass monopole moment $M$
\cite{BoR66}--\cite{Bo74}, \cite{CTJNew68,CH71,S90,BD88,BD92,BS93}.  A
signature of the presence of wave tails in the gravitational signals
received from inspiralling compact binaries should be detectable by VIRGO
and LIGO \cite{BSat94,BSat95}.

In order to determine more explicitly the relations (\ref{eq:27}) one must
implement to nonlinear order the coordinate transformation from harmonic to
radiative coordinates. This has been done to quadratic order in Section II
of \cite{BD92}. To see how this works let us first consider the leading-order
$1/r$ part when $r \to \infty$ (with $t-r/c$ fixed) of the linearized
metric $h^{\mu\nu}_1$ given by (\ref{eq:13})-(\ref{eq:14}).  One can write
\begin{equation}
  h^{\mu\nu}_1 = {1\over r} \left\{ -{4M\over c^2} \delta^\mu_0 \delta^\nu_0
  + z^{\mu\nu} (t-r/c,{\bf n}) \right\} + O \left( {1\over r^2} \right)\ ,
 \label{eq:29}
\end{equation}
where we have isolated the static term depending on $M$ from the non-static
contribution $z^{\mu\nu}$ given by
\begin{eqnarray}
 z^{00}(t,{\bf n})&=&- 4 \sum_{\ell\geq 2}\,{n_L \over c^{\ell +2}\ell !}
     \ M^{(\ell)}_L(t)\ , \nonumber \\ 
 z^{0i}(t,{\bf n})&=&- 4\sum_{\ell\geq 2}\,{n_{L-1}\over c^{\ell+2}\ell !}
       \ M^{(\ell)}_{iL-1} (t) \nonumber \\
   && + 4 \sum_{\ell \geq 2}\,{\ell \over c^{\ell +3}(\ell +1)!}
       \varepsilon_{iab}n_{aL-1} S^{(\ell)}_{\!bL-1}(t) ,
       \nonumber \\ 
 z^{ij}(t,{\bf n})&=&- 4\sum_{\ell\geq 2}\,{n_{L-2}\over c^{\ell +2}\ell !}
  \,M^{(\ell)}_{ijL-2}(t) \nonumber \\
   &&+ 8 \sum_{\ell\geq 2}\,{\ell\over c^{\ell +3}(\ell +1)!}\,
    n_{aL-2}\ \varepsilon_{ab(i}
  S^{(\ell)}_{\!j)bL-2}(t)\ .
     \label{eq:30}
\end{eqnarray}
Following the construction of the exterior field in Section 3.1 we insert
(\ref{eq:29}) into the quadratically nonlinear source (\ref{eq:9}) and arrive
at the equation to be satisfied by $h^{\mu\nu}_2$ at the leading-order in
$1/r$,
\begin{equation}
 \Box h^{\mu\nu}_2 = {1\over r^2} \left\{ {4M\over c^4} z^{(2)\mu\nu}
  + {k^\mu k^\nu\over c^2} \Pi \right\} (t-r/c,{\bf n})
+ O\left( {1\over r^3} \right)\ , \label{eq:31}
\end{equation}
where $k^\mu =(1,{\bf n})$ denotes the outgoing null direction and where $\Pi$
is proportional to the power $dE_1/dtd \Omega$
(per unit of steradian) contained in the linearized wave $h_1$~:
\begin{equation}
  \Pi = {1\over 2} z^{(1)\mu\nu} z^{(1)}_{\mu\nu}
    - {1\over 4} z^{(1)\mu}_\mu z^{(1)\nu}_\nu = {16\pi \over Gc^3}
    {dE_1\over dtd\Omega} \ . \label{eq:32}
\end{equation}
Constructing the solution of (\ref{eq:31}) using the method of Section
3.1 necessitates knowing the retarded integral of a source with
multipolarity $\ell$, and behaving in $r$ like $1/r^2$. The relevant
formula is (\ref{eq:21}). Working out this formula to
leading-order in $1/r$ (or $\ln r/r$ in fact) we get \cite{BD92}
\begin{eqnarray}
 &&\Box^{-1}_R \left[ \hat n_L r^{-2} F \left( t -{r\over c} \right)\right]
  = {c\hat n_L\over 2r} \int^{+\infty}_0 d\,y F \left(t-{r\over c}-y\right)
\left[\ln\left({cy\over 2r}\right)+\sum^\ell_{i=1} {2\over i}\right]\nonumber\\
 &&\qquad\qquad\qquad\qquad\qquad\quad + O \left( {\ln r\over r^2} \right)\ ,
\label{eq:33}
\end{eqnarray}
showing explicitly how some logarithms of $r$ are generated at the quadratic
nonlinear level. All these logarithms can be get rid off by implementing the
coordinate transform of \cite{B87}. With this approximation one finds
\begin{equation}
  X^\mu = x^\mu + G\, \xi^\mu + G^2 \lambda^\mu + O(G^3)\ , \label{eq:34}
\end{equation}
where the vectors $\xi^\mu$ and $\lambda^\mu$ are given by
\begin{eqnarray}
 \xi^\mu &=&-{2M\over c^2} \delta^\mu_0 \ln \left( {r\over cb} \right)\ ,
  \nonumber \\ 
 \lambda^\mu &=&\Box^{-1}_R \left[ {k^\mu\over 2cr^2} \int^{t-r/c}_{-\infty}
  dv \, \Pi (v,{\bf n}) \right]\ . \label{eq:35}
\end{eqnarray}
To order $G$ the coordinate transformation (\ref{eq:34})-(\ref{eq:35}) implies
(\ref{eq:23}). Both the vectors $\xi^\mu$ and $\lambda^\mu$ correct for
the deviation of the curved light cones from the flat cones in harmonic
coordinates. After coordinate transformation the metric is of the Bondi-type
and its $1/R$ term at infinity is readily obtained. By comparison of the
result with the expression of the waveform (\ref{eq:24}) one obtains the
looked-for observable moments in terms of the moments $M_L,S_L$. To quadratic
order and keeping only the contributions of non-local (or
hereditary) integrals one finds \cite{BD92}
\begin{eqnarray}
  U_L(T) &=&M^{(\ell)}_L (T) + {2GM\over c^3} \int^{+\infty}_0 dy\, \ln
   \left( {y\over 2b}\right) M_L^{(\ell+2)} (T-y) \nonumber \\
   &&+{Gc^{\ell+1}\ell!\over 2(\ell+1)(\ell+2)} \int^T_{-\infty} dv\,
    \Pi_L (v) + G{\cal I}_L (T) + O(G^2)\ , \nonumber \\ 
  V_L(T) &=&S^{(\ell)}_L (T) + {2GM\over c^3} \int^{+\infty}_0 dy\, \ln
   \left( {y\over 2b}\right) S_L^{(\ell+2)} (T-y) \nonumber\\
     &&+ G{\cal J}_L (T) + O(G^2) \ , \label{eq:36}
\end{eqnarray}
where $\Pi_L$ denotes the STF coefficient in factor of $\hat n_L$ in the STF
spherical harmonics decomposition of $\Pi$ (namely $\Pi =\Sigma \hat n_L
\Pi_L$), and where we denote by ${\cal I}_L$ and ${\cal J}_L$ some
instantaneous (by opposition to hereditary) contributions which are not
controlled at this stage (but see below).

The expressions (\ref{eq:36}) display two different types of non-local
integrals. The first type, arising in both $U_L$ and $V_L$ and having
$M$ in factor, represents the contribution of the backscattering
of the waves emitted in the past onto the static background curvature
associated with the mass of the source (tail effect). The constant $b$ in the
logarithmic kernel of the tail integrals is the same as in (23) and
(\ref{eq:35}). The second type of non-local integral, arising only in $U_L$
(at this order), is due to the re-radiation of gravitational waves by the
distribution of stress-energy of linear waves. We shall refer to this
non-local integral as the non-linear memory integral. The qualitatively
different nature of the tail and non-linear memory effects is more clearly
understood when taking the limit $T\to +\infty$ corresponding to very late
times after the source has ceased to emit radiation (in the sense that the
$(\ell+1)$th time derivatives of the moments $M_L$ and $S_L$ tend to
zero). In this limit the tail integrals (as well as the instantaneous
terms ${\cal I}_L$ and ${\cal J}_L$) tend to zero while the memory
integral tends to a finite limit (hence its name)~:
\begin{equation}
  U_L(+\infty) =M^{(\ell)}_L (+\infty)
    + {Gc^{\ell+1}\ell!\over 2(\ell+1)(\ell+2)} \int^{+\infty}_{-\infty} dv\,
    \Pi_L (v) + O(G^2)\ . \label{eq:37}
\end{equation}
The nonlinear memory term was first noticed by Payne \cite{P83} in a
particular context. It appeared in the form (\ref{eq:36}) in the present
author's habilitation thesis \cite{B90} (and later in \cite{BD92}).
Then the memory effect was obtained by Christodoulou \cite{Chr91} using
asymptotic techniques, and Thorne \cite{Th92} pointed out that the
effect can be recovered simply by adding to his formulas for the linear
memory \cite{BTh87} the contribution of gravitons.  Wiseman and Will
\cite{WiW91} evaluated the term in the quadrupole approximation and
found a result equivalent to the ones of \cite{B90,BD92}.  The nonlinear
memory term is however not very important in the case of inspiralling
compact binaries as compared with the terms associated with tails (and
even tails of tails).

It can be readily verified that in a post-Newtonian expansion $\varepsilon \to
0$ the tail integral present in the observable moment $U_L$ dominates the
memory integral (see section III.A in \cite{BD92}). Indeed the tail arises
dominantly at order 1.5PN (this is clear from the factor $c^{-3}$ in
front of this term), while the nonlinear memory term is smaller, being
at least of order 2.5PN.  For instance, one finds that in the
post-Newtonian expansion $\varepsilon \to 0$, the mass quadrupole
observable moment $U_{ij}$ (which is the most important moment to
obtain) reads, up to 1.5PN order, as
\begin{equation}
  U_{ij} (T)= M^{(2)}_{ij} (T) + {2GM\over c^3} \int^{+\infty}_0 dy \left[
   \ln \left( {y\over 2b}\right) + {11\over 12} \right] M^{(4)}_{ij}
  (T-y) + O(\varepsilon^5)\ . \label{eq:37'}
\end{equation}
The constant 11/12 in the integral is actually in factor of an instantaneous
term (depending only on $T$ through the third time derivative
$M^{(3)}_{ij} (T)$). It is computed in the appendix~B of \cite{BD92}.

Pushing the post-Newtonian expansion beyond the 2.5PN order controlled in
(\ref{eq:36}) requires investigating some cubic nonlinearities in the radiation
field. To 3PN order there appears an effect due to the ``tails of tails'',
which are the tails generated by backscattering of the (quadratically
nonlinear) tails themselves onto the background curvature associated
with $M$. This effect corresponds to the trilinear interaction between
two mass monopoles $M$ and (say) the quadrupole $M_{ij}$ describing the linear
radiation. The dominant tails of tails are computed in \cite{B96tail}.
Let us give now the expression of the observable mass quadrupole moment
up to the 3PN order ($<ij>$ denotes the STF projection)~:
\begin{eqnarray}
  U_{ij}(T) &=&M^{(2)}_{ij} (T) + 2{GM\over c^3} \int^{+\infty}_{0} dy
  \left[ \ln \left( {y\over 2b}\right) + {11\over 12} \right] M^{(4)}_{ij}
  (T-y) \nonumber \\
&+&{G\over c^5} \Biggl\{ - {2\over 7} \int^T_{-\infty} dv\, M^{(3)}_{k<i}
  (v) M^{(3)}_{j>k} (v) +\alpha M^{(3)}_{k<i} (T) M^{(2)}_{j>k} (T) \nonumber\\
&+&\beta M^{(4)}_{k<i}(T)M^{(1)}_{j>k}(T)+\gamma M^{(5)}_{k<i}(T) M_{j>k}(T)
  + \lambda S_k M^{(4)}_{m<i} (T) \varepsilon_{j>km} \Biggr\} \nonumber\\
&+&2\left( {GM\over c^3}\right)^2 \int^{+\infty}_0 dy \left[ \ln^2
   \left( {y\over 2b}\right) + {57\over 70} \ln \left( {y\over 2b}\right)
 + \sigma \right] M^{(5)}_{ij} (T-y) \nonumber \\
&+&O(\varepsilon^7)\ . \label{eq:38}
\end{eqnarray}
The various terms are easily recognizable on this expression. Besides the
dominant tail
term of order $\varepsilon^3 \approx c^{-3}$, there is the lowest order
contribution due to the nonlinear memory which is the integral appearing at
order $\varepsilon^5$ (it depends quadratically on the quadrupole moment
\cite{B90,WiW91,BD92}), and there is the tail of tail term which is
given by the last integral of order $\varepsilon^6$ (one assumes that $b$
enters (\ref{eq:17}) in the combination $(r/cb)^B)$. The $\alpha$, $\beta$,
$\gamma$, $\lambda$, and $\sigma$ are some purely numerical constants in factor
of instantaneous terms (like the 11/12), and which are computed in
\cite{B96tail}. Note that the precise numerical value of these constants is
not very important physically (they partly reflect our
use of harmonic coordinates), but their computation using the formulas
of Section~3.2 (and other formulas) is in general long and tedious.

The observable multipole moments other than the quadrupole admit some
expressions similar to (\ref{eq:38}), but thanks to the fact that the
multipolarity of a contribution scales with $1/c$ in the waveform (\ref{eq:24})
and with $1/c^2$ in the energy loss (\ref{eq:25}), the needed accuracy
for these moments is less than for the quadrupole. We quote here the
expressions of the mass octupole $U_{ijk}$ and current quadrupole $U_{ij}$
with an accuracy sufficient to include the dominant tails only~:
\begin{eqnarray}
 U_{ijk} (T) &=& M^{(3)}_{ijk} (T) + 2{GM\over c^3} \int^{+\infty}_0 dy
  \left[ \ln \left( {y\over 2b}\right) + {97\over 60}\right]
  M^{(5)}_{ijk} (T-y) +O(\varepsilon^5) , \nonumber \\ 
 V_{ij} (T) &=& S^{(2)}_{ij} (T) + 2{GM\over c^3} \int^{+\infty}_0 \! dy
  \left[ \ln \left( {y\over 2b}\right) + {7\over 6}\right]
  S^{(4)}_{ij} (T-y) +O(\varepsilon^5) .\nonumber \\ \label{eq:39}
\end{eqnarray}
The computation of the constants 97/60 and 7/6 can be found in Appendix C
of \cite{B95}. We assume in (\ref{eq:39}) that our coordinate system is
mass-centered, $M_i=0$.

\section{GENERATION OF GRAVITATIONAL WAVES} \label{sec:4}

\subsection{Post-Newtonian iteration of the field in $D_i$}  \label{sec:4.1}

We now come to grips with the last step of the present approach, which is to
find the expressions of the multipole moments $M_L$ and $S_L$ as explicit
integrals over the distribution of stress-energy of the matter fields (and the
gravitational field) in the source. This problem is evidently crucial in this
approach since without a satisfactory solution the formulas
(\ref{eq:36})-(\ref{eq:39}) giving the observables
measured far from the source would remain devoided of
meaning, and the precise interpretation of the gravitational signals
received by the future detectors would be impossible.

To be consistent with the formula (\ref{eq:38}) for instance, we should obtain the
mass quadrupole moment $M_{ij}$ with a precision corresponding to the  3PN
order, thereby generalizing the Newtonian relation (\ref{eq:15}) to take into
account all the relativistic corrections up to the very high order
$\varepsilon^6$. As we shall see the 3PN precision in this moment is not
available presently, but $M_{ij}$ is known with the precision which precedes
immediately, namely the 2.5PN or $\varepsilon^5$ precision \cite{B95,B95pn}.

The non-vacuum Einstein field equations in harmonic coordinates read
\begin{eqnarray}
 \Box h^{\mu\nu} &=&{16\pi G\over c^4} {\cal T}^{\mu\nu} +\Lambda^{\mu\nu} (h)
  \ , \nonumber \\ 
 \partial_\nu h^{\mu\nu} &=&0 \ , \label{eq:40}
\end{eqnarray}
where ${\cal T}^{\mu\nu} \equiv |g| T^{\mu\nu}$ represents the stress-energy
distribution of the matter fields (with $T^{\mu\nu}$ the usual stress-energy
tensor), and where the other notations are defined in Section 3.1.
In the near-zone $D_i$ the field equations can be solved by means of a direct
post-Newtonian iteration when $\varepsilon \to 0$. We shall perform it
without making any assumption concerning the stress-energy distribution
${\cal T}^{\mu\nu}$ (it could be for instance that of a perfect fluid,
or it could describe point particles without interactions), except that
it should have a compact support and correspond to a source whose
parameter $\varepsilon$ is of the order 0.3 (say) at most.

It is useful to introduce as basic variables describing the source some
densities of mass $\sigma$, current $\sigma_i$ and stresses $\sigma_{ij}$
defined in terms of the contravariant components of the stress-energy
tensor $T^{\mu\nu}$ (with $T^{ii} =\Sigma \delta_{ij} T^{ij})$~:
\begin{eqnarray}
 \sigma &=& {T^{00} + T^{ii} \over c^2}\ , \nonumber  \\ 
 \sigma_i &=& {T^{0i} \over c}\ , \nonumber \\ 
 \sigma_{ij} &=& T^{ij} \ . \label{eq:41}
\end{eqnarray}
In particular $\sigma$ is related to the Tolman mass density valid for
stationary systems \cite{LL0}.  Here the powers of $c^{-1}$ are such that
these densities have a finite non-zero limit when $\varepsilon \to 0$
($T^{\mu\nu}$ has the dimension of an energy density). The usefulness of
these definitions lies in the fact that they simplify appreciably the
first post-Newtonian approximation (this was pointed out in Section II of
\cite{BD89}), and therefore, as we shall see, the subsequent approximations
built on it. See the lecture of T.~Damour in this volume
for the derivation of the 1PN approximation using the source densities
(\ref{eq:41}), and for applications to the relativistic N-body problem.

{}From the densities (\ref{eq:41}) we introduce next the following {\it
retarded} potentials
\begin{eqnarray}
 V &=& -4\pi\,G\,\Box^{-1}_R \sigma \ , \nonumber \\ 
 V_i &=& -4\pi\,G\,\Box^{-1}_R \sigma_i \ ,\nonumber \\ 
 W_{ij} &=& -4\pi\,G\,\Box^{-1}_R \left[ \sigma_{ij} + {1\over 4\pi G}
 \left( \partial_i V \partial_j V - {1\over 2} \delta_{ij} \partial_k V
  \partial_k V \right) \right] \ ,  \label{eq:42}
\end{eqnarray}
where $\Box^{-1}_R$ is the flat retarded d'Alembertian operator (\ref{eq:16}).
The scalar and vectorial potentials $V$ and $V_i$ reduce in the limit
$\varepsilon \to 0$ to the Newtonian potential $U$ and the usual
``gravitomagnetic'' potential $U_i$. The tensorial potential $W_{ij}$
is more complicated but is generated in the limit $\varepsilon \to 0$ by the
sum of the matter stress density in the source and the stresses associated
with the Newtonian gravitational field itself (the gravitational stresses
are of the same order as the matter stresses when $\varepsilon \to 0$).
The sources of the potentials $V$ and $V_i$ are of compact support, while
the source of $W_{ij}$ is non compact. Thanks to the equations of continuity
and of motion for the source densities (\ref{eq:41}) these potentials
satisfy some simple differential identities to Newtonian order~:
$\partial_t V+\partial_i V_i= O(\varepsilon^2)$ and $\partial_t V_i
+\partial_j W_{ij} =O(\varepsilon^2)$.  Beware of the fact that these
potentials are different from the external potentials introduced in
(\ref{eq:13})-(\ref{eq:14}).

To Newtonian order one has $h^{00} = -4V/c^2 + O(\varepsilon^4)$, $h^{0i} =
O(\varepsilon^3)$ and $h^{ij} =O(\varepsilon^4)$, that we can substitute into
the right-hand-side of the wave equation in (\ref{eq:40}), where with this
approximation the nonlinear term $\Lambda^{\mu\nu} (h)$ can be replaced by its
quadratic nonlinear piece given by (\ref{eq:9}). Applying $\Box^{-1}_R$ to the
result (without any finite part of course because the source is regular
in $D_i$), and neglecting consistently all the small terms, one obtains
the first post-Newtonian approximation in the form \cite{B95}
\begin{eqnarray}
 h^{00}_{\rm in} &=& -{4\over c^2} V +{4\over c^4} (W_{ii} - 2V^2)
  +O(\varepsilon^6)\ ,   \nonumber \\ 
 h^{0i}_{\rm in} &=& -{4\over c^3} V_i
 +O(\varepsilon^5)\ ,\nonumber \\ 
 h^{ij}_{\rm in} &=& -{4\over c^4} W_{ij} +O(\varepsilon^6)\ .\label{eq:43}
\end{eqnarray}
Actually this approximation is slightly more accurate than the 1PN
approximation because it takes into account the terms of
order $\varepsilon^4$ not only in the $00$ component of the metric, but also
in its $ij$ components. We indicate by the subscript ``in'' that the metric is
the inner metric valid in $D_i$, which differs by a coordinate transformation
from the exterior metric of Section 3 (see below). Substituting the expressions
(\ref{eq:43}) back into the right-hand-side of the field equation in
(\ref{eq:40}), where now the nonlinear term $\Lambda^{\mu\nu} (h)$ should
include besides the quadratically nonlinear piece (\ref{eq:9}) a cubically
nonlinear piece (given in Section I.B of \cite{B95}), neglecting consistently
the higher order terms and then inverting the result by means of
$\Box^{-1}_R$, one gets
\begin{equation}
  h^{\mu\nu}_{\rm in} = \Box^{-1}_R \left[ {16\pi G\over c^4}
     \overline{\cal T}^{\mu\nu} + \overline \Lambda^{\mu\nu} (V,W) \right]
    + O(\varepsilon^8, \varepsilon^7, \varepsilon^8)\ , \label{eq:44}
\end{equation}
where the remainder means that the neglected terms are $O(\varepsilon^8)$,
$O(\varepsilon^7)$ and $O(\varepsilon^8)$ in the components $h^{00}_{\rm in}$, 
$h^{0i}_{\rm in}$ and $h^{ij}_{\rm in}$, respectively. The quantities 
$\overline{\cal T}^{\mu\nu}$ and $\overline\Lambda^{\mu\nu} (V,W)$ are given by
the corresponding quantities in the right-hand-side of (\ref{eq:40}), but
where one retains only the terms in the post-Newtonian expansion up to the
accuracy of the remainder in (\ref{eq:44}). These are given as explicit
combinations of derivatives of $V$, $V_i$ and $W_{ij}$.  For instance
the effective nonlinear source $\overline\Lambda^{\mu\nu} (V,W)$ reads
\begin{eqnarray}
 \overline\Lambda^{00}(V,W) &=& - {14\over c^4} \partial_k V
 \partial_k V + {16\over c^6} \biggl\{ - V \partial^2_t V - 2V_k \partial_t
 \partial_k V \nonumber \\
 && \qquad - W_{km} \partial_{km}^2 V + {5\over 8} (\partial_t V)^2
 + {1\over 2} \partial_k V_m (\partial_k V_m +3\partial_m V_k) \nonumber\\
 && \qquad + \partial_k V\partial_t V_k + 2\partial_k V\partial_k W_{mm}
  - {7\over 2} V\partial_k V\partial_k V \biggr\} \ , \nonumber\\ 
 \overline\Lambda^{0i}(V,W) &=&  {16\over c^5} \left\{ \partial_k V
 (\partial_i V_k - \partial_k V_i) + {3\over 4} \partial_t V \partial_i V
 \right\} \ , \nonumber \\ 
 \overline\Lambda^{ij}(V,W) &=& {4\over c^4}\left\{ \partial_i V
 \partial_j V -{1\over 2} \delta_{ij} \partial_k V \partial_k V \right\}
  + {16\over c^6} \biggl\{ 2 \partial_{(i} V\partial_t V_{j)}
  - \partial_i V_k \partial_j V_k  \nonumber \\
 && \qquad - \partial_k V_i \partial_k V_j
  + 2 \partial_{(i} V_k \partial_k V_{j)} - {3\over 8} \delta_{ij}
    (\partial_t V)^2 - \delta_{ij} \partial_k V \partial_tV_k \nonumber \\
 && \qquad + {1\over 2} \delta_{ij}\partial_k V_m (\partial_k V_m
   - \partial_m V_k) \biggr\} \ . \label{eq:45}
\end{eqnarray}
See the section II.A of \cite{B95} for the derivation of these formulas.

\subsection{The multipole moments as integrals over the source} \label{sec:4.2}

The expressions of the multipole moments $M_L$ and $S_L$ will follow from the
comparison between the inner metric (\ref{eq:44})-(\ref{eq:45}) valid in $D_i$
and the {\it re-expansion} when $\varepsilon \to 0$ of the exterior metric
valid in $D_e$.  The comparison (and in fact matching) is done in the
overlapping region $D_e \cap D_i$ which always exists for slowly-moving
sources. This ``matching'' region is where one can simultaneously expand
the field when $\varepsilon \to 0$ and decompose it into multipole moments.
The matching requirement is simply that the two inner and outer post-Newtonian
asymptotic expansions should be (term by term) identical in $D_i \cap
D_e$, after the performing of a suitable coordinate transformation.
Let this coordinate transformation be
\begin{equation}
 x^\mu_{\rm ext} = x^\mu _{\rm in} + \varphi^\mu (x_{\rm in})\ , \label{eq:46}
\end{equation}
where $x^\mu_{\rm ext}$ and $ x^\mu_{\rm in}$ denote the coordinates which were 
used in $D_e$ and $D_i$ respectively (with a slight change of notation
with respect to Section~3).

Recall that the exterior metric is constructed starting with the linearized
metric $h^{\mu\nu}_1$ whose components are defined by some external potentials
$V^{\rm ext}$, $V^{\rm ext}_i$ and $V^{\rm ext}_{ij}$ given explicitly as
some infinite multipole expansions parametrized by $M_L$ and $S_L$ (see
(\ref{eq:13})-(\ref{eq:14})). The problem is to relate via the matching
requirement (\ref{eq:46}) these exterior potentials to the inner potentials
$V$, $V_i$ and $W_{ij}$ used in $D_i$. In fact we do so first with some
intermediate accuracy.  The method is to look for a {\it numerical} equality
in $D_i \cap D_e$ between exterior and inner potentials, and then transform
this equality into a {\it matching} equation (i.e.  an equation relating
two mathematical expressions of the same nature) by replacing the inner
potentials by their {\it multipole expansions} in $D_e$.  The matching
equation is valid formally ``everywhere", and it can be used to get the
functional relationships linking the multipole moments to the source.

For instance it is found that the matching equations relating the potentials
$V^{\rm ext}$ and $V^{\rm ext}_i$ to the multipole expansions ${\cal M}(V)$
and ${\cal M}(V_i)$ of the compact-support inner potentials $V$
and $V_i$ (where ${\cal M}$ refers to the multipole expansion) read as
\begin{eqnarray}
  V^{\rm ext} &=& {\cal M} (V) + c\partial_t \varphi^0 + O(\varepsilon^4)\ ,
   \nonumber \\ 
  V^{\rm ext}_i &=& {\cal M} (V_i) - {c^3\over 4} \partial_i \varphi^0
      + O(\varepsilon^2) \ ,  \label{eq:47b}
\end{eqnarray}
where $\varphi^0$ denotes the zero component of the gauge transformation vector
in (\ref{eq:46}). The equation for $V^{\rm ext}$ is valid to first
post-Newtonian order, but the equation for $V^{\rm ext}_i$ is valid only to
Newtonian order (it can be proved that $\varphi^0$ is of order
$\varepsilon^3$). The matching equations (\ref{eq:47b}) were obtained
in Section II of \cite{BD89}.  The multipole expansion of a retarded
potential with {\it compact-support} source is known from the work of
Campbell, Macek and Morgan \cite{CMM77}.  It has been re-calculated in
the appendix B of \cite{BD89} using STF spherical harmonics.
Using this formula Damour and Iyer \cite{DI91b} improved much the work
\cite{CMM77} dealing with the case of linearized gravity.
The result (with obvious notation for the index $\mu$) is
\begin{equation}
 {\cal M}(V_\mu) = G \sum_{\ell\geq 0} {(-)^\ell\over \ell !}
 \partial_L \left[ {1\over r} {\cal V}_\mu^L \left( t-{r\over c}\right)\right]
  \ , \label{eq:48}
\end{equation}
where the multipole moments which parametrize the expansion are given by
\begin{equation}
 {\cal V}_\mu^L(t) = \int d^3{\bf x}\,\hat x_L \int^1_{-1} dz\, \delta_\ell
 (z)\sigma_\mu ({\bf x},t+z |{\bf x}|/c)\ .\label{eq:49}
\end{equation}
We denote by $\hat x_L$ the STF part of the product of $\ell$ spatial
positions $x_L=x_{i_1} \cdots x_{i_\ell}$. The function $\delta_{\ell} (z)$
takes into account the physical delays due to the propagation of the waves 
inside the compact-support source. It is given by
\begin{equation}
\delta_\ell (z) ={(2\ell +1)!!\over 2^{\ell+1}\ell !} (1-z^2)^\ell\quad ;
 \qquad \quad \int^1_{-1} dz\, \delta_\ell (z) =1\  \label{eq:50}
\end{equation}
($\delta_\ell$ can be simply related via the Rodrigues formula to the
Legendre polynomial of order $\ell$).
With the two matching equations  (\ref{eq:47b}) and the explicit
formulas (\ref{eq:49})-(\ref{eq:50}), it is a simple question of algebraic
manipulation (using STF techniques), and analytic post-Newtonian
expansion (using the formula (B.14) of \cite{BD89}), to deduce the
multipole moments $M_L$ and $S_L$ entering the left-hand-sides of
(\ref{eq:47b}).  In particular the mass moment $M_L$ is
obtained with post-Newtonian accuracy as
\begin{equation}
  M_L (t)=\!\int\! d^3 {\bf x} \Biggl\{ \hat x_L \sigma
  +{|{\bf x}|^2 \hat x_L\over 2c^2 (2\ell+3)} \partial_t^2 \sigma
  -{4(2\ell+1)\hat x_{iL}\over c^2(\ell+1)(2\ell+3)} \partial_t \sigma_i
   \Biggr\} + O(\varepsilon^4)\ . \label{eq:51}
\end{equation}
This expression generalizes to 1PN order the result (\ref{eq:15}) of the
Newtonian formalism. The striking feature about this expression is that the
integrand has a compact support, althought it includes the contribution of
the gravitational field to 1PN order, which is expected to extend over the
whole three-dimensional space. However, the gravitational contribution
to 1PN order turns out to be entirely contained into the mass density $\sigma$.
As the expression (\ref{eq:51}) has been derived rigorously within the
post-Newtonian framework, it can also be said (when combined with the
previous results in $D_w$) to bring a satisfactory proof of the
lowest-order quadrupole formula itself.

To the next order (2PN) things are more delicate precisely because of the
explicit contributions of the gravitational field which come into play,
making the expressions of the mass multipole moments a priori non compact.
This problem has been solved in \cite{B95}, where it was found necessary
to obtain first the expression of the multipole expansion ${\cal M}(W_{ij})$
of the inner potential $W_{ij}$ defined in (\ref{eq:42}), and whose source we
recall is non-compact. Thus the formulas (\ref{eq:49})-(\ref{eq:50})
cannot be employed in this case, but instead we have obtained (in
Section III.C of \cite{B95})
\begin{eqnarray}
&&{\cal M}(W_{ij})=G \sum_{\ell\geq 0} {(-)^\ell\over \ell !}
 \partial_L \left[{1\over r}{\cal W}^L_{ij}\left( t-{r\over c}\right)\right]
  \nonumber\\
 &&\qquad+{\rm FP}_{B=0}\Box^{-1}_R\left[ r^B \left( -\partial_i {\cal M}(V)
  \partial_j {\cal M}(V) + {1\over 2} \delta_{ij} \partial_k {\cal M}(V)
  \partial_k {\cal M}(V) \right) \right] , \nonumber \\ \label{eq:52}
\end{eqnarray}
where the multipole moments are given by
\begin{eqnarray}
 &&{\cal W}_{ij}^L(t) = {\rm FP}_{B=0} \int d^3 {\bf x}\,
  |{\bf x}|^B \hat x_L\qquad\qquad\qquad \nonumber \\
 &&\qquad\times\int^1_{-1} dz\, \delta_\ell (z)
\left[ \sigma_{ij} +{1\over 4\pi G}
   \left(\partial_i V\partial_j V
   - {1\over 2} \delta_{ij} \partial_k V\partial_k
   V\right)\right] ({\bf x}, t+z |{\bf x}|/ c)\ . \nonumber \\
\label{eq:53}
\end{eqnarray}
The first term in (\ref{eq:52}) together with the moments (\ref{eq:53})
look like exactly what we would obtain by using blindly the formulas
(\ref{eq:49})-(\ref{eq:50}), even though the potential is non compact.
However there is an important difference which is the presence in the moments
(\ref{eq:53}) of the finite part procedure which was used in the definition
of the exterior metric (see (\ref{eq:17})). The role of this finite part is to
deal with the infinite support of the integral (\ref{eq:53}), which would be
divergent otherwise (because of the presence in the integrand of $\hat x_L$
behaving like $|{\bf x}|^\ell$ at spatial infinity). Notice that the finite
part is conveyed all the way through the analysis, and is seen {\it a
posteriori} to be in fact necessary to ensure the convergence of the
integrals. Thus no {\it ad hoc} prescription has been used to obtain the
moments (\ref{eq:53}) in a well-defined form mathematically, even in the case
of a non-compact supported source (see the proof in \cite{B95}). As for the
second term in (\ref{eq:52}), it ensures as is also seen a posteriori
that ${\cal M}(W_{ij})$ satisfies the
correct wave equation (deduced from (\ref{eq:42})) outside the source. This
second term involves the multipole expansion ${\cal M}(V)$ of $V$ in
conformity with the retarded integral operator ${\rm FP}_{B=0}
\Box^{-1}_R r^B$ which is defined only when acting on multipole expansions.

With this sub-problem of obtaining ${\cal M}(W_{ij})$ solved we can write
down similarly to (\ref{eq:47b}) a matching equation relating the external
potential $V^{\rm ext}_{ij}$ to the multipole expansion ${\cal M}(W_{ij})$.
Together with (\ref{eq:47b}) this permits us to relate the nonlinear source in
the exterior field, which to this post-Newtonian order is shown to be
exactly $\overline\Lambda^{\mu\nu} (V^{\rm ext},W^{\rm ext})$, where
$\overline\Lambda^{\mu\nu}$ is given by (\ref{eq:45}) and where $W^{\rm ext}
_{ij}$ is defined from $V^{\rm ext}_{ij}$, to the
multipole expansion of the nonlinear source, that is
$\overline\Lambda^{\mu\nu} ({\cal M}(V),{\cal M}(W)$), modulo the
terms associated with the coordinate transformation. This relation being
established one can write a matching equation valid to higher post-Newtonian
order. The nonlinear part of the coordinate transformation is found to cancel
out, so that it remains simply a linear gauge transformation associated with
the vector $\varphi^\mu$. Finally a reasoning similar to the one followed to
obtain (\ref{eq:52})-(\ref{eq:53}) leads to the linear piece $h^{\mu\nu}_1$
of the external field (which is a functional of the algorithmic moments $M_L$
and $S_L$) in the form of the explicit multipole expansion
\begin{equation}
 Gh^{\mu\nu}_{1} [M_L,S_L] = -{4G\over c^4} \sum_{\ell\geq 0}
 {(-)^\ell\over \ell !} \partial_L \left[ {1\over r}\,
  {\cal F}^{\mu\nu}_L \left( t-{r\over c}\right)\right] +
  \partial\varphi^{\mu\nu} + O(\varepsilon^7)\ , \label{eq:54}
\end{equation}
where $\partial\varphi^{\mu\nu} =\partial^\mu\varphi^\nu +\partial^\nu
\varphi^\mu -\eta^{\mu\nu} \partial_\lambda \varphi^\lambda$ is the
linear gauge transformation, which depends on some (reducible) multipole
moments given by
\begin{equation}
 {\cal F}^{\mu\nu}_L(t) = {\rm FP}_{B=0} \int
  d^3 {\bf x}|{\bf x}|^B \hat x_L \int^1_{-1} dz\, \delta_\ell (z)
  \overline\tau^{\mu\nu} \left( {\bf x}, t+z {|{\bf x}|/ c}\right)\ .
  \label{eq:55}
\end{equation}
The integrand of these multipole moments is nothing but the total
stress-energy momentum of the matter fields {\it and} the gravitational
field (with 2PN accuracy), namely
\begin{equation}
   \overline\tau^{\mu\nu} = \overline{\cal T} ^{\mu\nu} +
   {c^4\over 16\pi G} \overline\Lambda^{\mu\nu} (V, W)\ , \label{eq:55'}
\end{equation}
which is given, thanks (in particular) to (\ref{eq:45}), as a fully {\it
explicit} functional of the {\it inner} potentials $V$, $V_i$ and $W_{ij}$.
Like in (\ref{eq:53}) there is a Finite Part symbol in front of the
integral which is seen (a posteriori) to make the non-compact-support
multipole moments to be mathematically well-defined.

The result (\ref{eq:54})-(\ref{eq:55'}) (proved in Section III.D of \cite{B95})
is satisfying because one recovers finally the expression of the total
pseudotensor of the matter and gravitational fields in the source, which is
exactly the one which appears in the right-hand-side of the Einstein field
equation (\ref{eq:40}), except that here it is expanded when $\varepsilon\to 0$
up to 2PN order. Notice that although the range of integration extends up to
infinity, it is the post-Newtonian expansion valid in $D_i$ of the
pseudotensor which is to be used in (\ref{eq:55}). This is thanks to the
finite part procedure and the properties of analytic continuation
(see \cite{B95} for the reason why this works).

The {\it form} of the result (\ref{eq:54})--(\ref{eq:55'}) is therefore
identical to what we would obtain by writing from the field equation
(\ref{eq:40}) the solution directly as the retarded integral $\Box^{-1}_R$
of the total pseudotensor (this operation is perfectly licit), but then by
writing uncorrectly the multipole expansion of that solution applying
the formulas (\ref{eq:48})-(\ref{eq:50}). This second operation is illicit
because the formulas apply only to compact-supported sources. As a result the
multipole moments would be given by divergent integrals. This was the
approach initially followed by Epstein and Wagoner \cite{EW75} to 1PN order,
and then generalized formally by Thorne \cite{Th80} to all post-Newtonian
orders.  The error is not too severe to 1PN order because one actually
recovers the correct result by discarding the infinite surface terms
in the Epstein-Wagoner moments.  This has been proved using the
compact-support formula (\ref{eq:51}) (see the appendix A of
\cite{BD89}), and this has also been checked in applications to binary
systems \cite{BS89}.  However in higher-orders the neglect of infinite
surface terms in the Epstein-Wagoner moments becomes physically
uncorrect, because it is then equivalent to the neglect of physical
effects in the waveform, namely the tail and other nonlinear effects
computed in Section 3.4.  This is clear from the fact that by
(\ref{eq:54})-(\ref{eq:55'}) one sees that the multipole moments which
are given by (the finite part of) integrals over the total stress-energy
pseudotensor are simply the moments $M_L$ and $S_L$, which differ from
the observable moments $U_L$ and $V_L$ in the wave field
(\ref{eq:24}) by all the nonlinear effects in $D_w$.

Recently Will and Wiseman \cite{WWi95} have succeeded in curing the defects of
the Epstein-Wagoner approach. As a result they obtain a manifestly convergent
and finite procedure for calculating the gravitational radiation from isolated
systems to post-Newtonian order.

It still remains to compute from (\ref{eq:54})-(\ref{eq:55'}) the (irreducible)
moments $M_L$ and $S_L$, which means to decompose in the right-hand-side
of (\ref{eq:54}) the moments ${\cal F}^{\mu\nu}_L$ into irreducible STF
pieces.  Apart from the Finite Part symbol in front of the moments and
the post-Newtonian remainder in (\ref{eq:54}), this technical problem is the
same as in the case of linearized gravity in which case it has been solved
by Damour and Iyer \cite{DI91b}.  The mass multipole moment $M_L$ to 2PN
order is therefore deduced, and, after further transformations done in
Section IV.B of \cite{B95}, we end up with
\begin{eqnarray}
{M}_L(t)&=&{\rm FP}_{B=0}\int d^3 {\bf x}|{\bf x}|^B
 \biggl\{ \hat x_L \biggl[\sigma + {4\over c^4}(\sigma_{ii}U -\sigma P_{ii})
 \biggr] \nonumber\\
 &+&{|{\bf x}|^2\hat x_L\over 2c^2(2\ell+3)} \partial^2_t
\sigma-{4(2\ell+1)\hat x_{iL}\over c^2(\ell+1)(2\ell+3)} \partial_t
   \left[ \left( 1 +{4U\over c^2} \right) \sigma_i \right.\nonumber \\
&+&\left.{1\over\pi Gc^2} \left(\partial_k U[\partial_i U_k -\partial_k U_i]
  + {3\over 4} \partial_t U \partial_i U \right) \right] \nonumber \\
&+&{|{\bf x}|^4\hat x_L\over 8c^4(2\ell+3)(2\ell+5)} \partial^4_t\sigma
   - {2(2\ell+1)|{\bf x}|^2\hat x_{iL}\over c^4(\ell+1)(2\ell+3)(2\ell+5)}
   \partial^3_t\sigma_i  \nonumber\\
&+&{2(2\ell+1)\over c^4(\ell+1)(\ell+2)(2\ell+5)} \hat x_{ijL}
   \partial_t^2 \left[ \sigma_{ij} + {1\over 4\pi G} \partial_i U
   \partial_j U \right] \nonumber\\
&+&{1\over \pi Gc^4}\hat x_L \biggl[ -P_{ij} \partial_{ij}^2 U
  - 2 U_i\partial_t\partial_i U + 2\partial_i U_j \partial_j U_i \nonumber \\
&&\qquad\quad- {3\over 2}(\partial_t U)^2 -U\partial_t^2 U\biggr] \biggr\}
  + O(\varepsilon^5) \ . \label{eq:56}
\end{eqnarray}
The potentials $U$, $U_i$ and $P_{ij}$ are the Newtonian-like potentials
to which tend the potentials $V$, $V_i$ and $W_{ij}$ in the limit $\varepsilon
\to 0$. As we see there are many explicit contributions to 2PN order of the
non-compact-supported distribution of the gravitational field. The expression
(\ref{eq:56}) has been generalized recently to include the next
2.5PN correction \cite{B95pn}.
Similarly one deduces also the expression of the current multipole moment
$S_L$, but with 1PN accuracy only~:
\begin{eqnarray}
{S}_L(t) &=&{\rm FP}_{B=0}\,\varepsilon_{ab<i_\ell}
   \int d^3 {\bf x}|{\bf x}|^B
   \biggl\{ \hat x_{L-1>a}\left( 1+{4\over c^2}U\right) \sigma_b
  + {|{\bf x}|^2\hat x_{L-1>a}\over
   2c^2 (2\ell+3)} \partial^2_t \sigma_b \nonumber \\
&&\qquad + {1\over \pi Gc^2} \hat x_{L-1>a}
  \left[ \partial_k U(\partial_b U_k -\partial_k U_b)
  + {3\over 4} \partial_t U \partial_b U \right] \nonumber \\
&&\qquad - {(2\ell +1)\hat x_{L-1>ac}\over c^2(\ell+2)(2\ell+3)}
 \partial_{t}\left[ \sigma_{bc} +{1\over 4\pi G}  \partial_b U\partial_c
   U \right] \biggr\} +O(\varepsilon^4)\ .  \label{eq:57}
\end{eqnarray}
The first correct expression for the 1PN current multipole moment was in fact
obtained earlier by Damour and Iyer \cite{DI91a}, making extensive use of
distributional kernels to deal with the quadratic nonlinearities of the
gravitational field. Their result has the advantage over (\ref{eq:57})
to explicitly express the 1PN current moments $S_L$ in terms of
compact-support integrals.
Its extension to higher orders would require to generalize the quadratic
kernels they introduce to cubic and higher kernels. [In a sense this
generalization has been performed in Section~III.C of \cite{BDI95},
where the result (58) has been transformed in an explicitly
compact-support integral.]
See Section IV.B of \cite{B95} for the complete
equivalence of (\ref{eq:57}) with the result of \cite{DI91a}.

\section{GRAVITATIONAL RADIATION REACTION EFFECTS} \label{sec:5}

On physical grounds one expects that it should be possible to any
post-Newton\-ian order to interpret the results concerning the outgoing
radiation field in $D_w$ (and the energy, linear momentum and angular
momentum it carries) as due to some radiation reaction force acting {\it
locally} inside the source (in $D_i$).  Indeed this is known to be correct
at the Newtonian level \cite{CE70}, \cite{AD75}--\cite{A87}. In Section~2.2
is reviewed the Burke and Thorne \cite{Bu69}--\cite{Bu71} expression for
the reaction force at this level. Other expressions of the
force in other coordinate systems were obtained by several authors, notably
Chandrasekhar and Esposito \cite{CE70} (see \cite{S83} for a discussion).
Recall that the (Newtonian) radiation reaction force is directly
responsible for the observed acceleration of the orbital motion of the
binary pulsar PSR~1913+16 \cite{TFMc79,D83b}.
 
Unfortunately the problem of radiation reaction onto the source is not so well
understood as the problem of the generation of radiation in the regions far
from the source (sections 3-4). Here we shall report the results of the
extension to first post-Newtonian order (and even 1.5PN order) of the
lowest-order Burke and Thorne radiation reaction force (6). These results
confirm to this order the physical expectation that all effects of the reaction
force are contained in the fluxes of energy and momenta carried out by the
waves at infinity, but clearly some more work should be done on this problem.

The novel feature when one goes from the Newtonian reaction force (6) to the 
first post-Newtonian one is that the reaction potential is no longer composed
of a single scalar (depending on the mass-type multipole moments), but involves
also a vectorial component depending on the variations of the
{\it current}-type multipole moments. The vectorial component of the
reaction could be important in some astrophysical situations like rotating
neutron stars undergoing gravitational instabilities. It was first noticed in
the physically restricting case where the dominant quadrupolar radiation from
the source is suppressed \cite{BD84}.

{}From the general case investigated in \cite{B93,B95reac} it follows in
particular that the expressions of the scalar and vectorial
reactive potentials which generalize to 1PN order the Burke-Thorne
potential (6) are given by
\begin{eqnarray}
  V^{\rm reac} ({\bf x},t) &=& - {G\over 5c^5} x_{ij} M^{(5)}_{ij} (t) +
  {G\over c^7} \left[ {1\over 189} x_{ijk} M^{(7)}_{ijk} (t) - {1\over 70}
  |{\bf x}|^2 x_{ij} M^{(7)}_{ij} (t) \right] \nonumber \\
   &&+ O (\varepsilon^8)   \ , \nonumber \\
  V_i^{\rm reac} ({\bf x},t) &=& {G\over c^5} \left[ {1\over 21} \hat
   x_{ijk} M^{(6)}_{jk} (t) - {4\over 45} \varepsilon_{ijk} x_{jm}
  S^{(5)}_{km} (t) \right] + O(\varepsilon^6)\ .\label{eq:58b}
\end{eqnarray}
One recognizes in the first term of $V^{\rm reac}$ the Burke-Thorne scalar
potential for the mass quadrupole moment $M_{ij}$. The vector potential
$V^{\rm reac}_i$ depends notably on the current quadrupole moment $S_{ij}$.
It is crucial that $M_{ij}$ in the Burke-Thorne term should be given with
relative 1PN precision in order to be consistent with the accuracy of the
approximation. The correct expression for that moment is (\ref{eq:51}) above
(taken for $\ell=2$). The other moments in (\ref{eq:58b}) take their usual
Newtonian forms at this approximation. Notice that the expressions
(\ref{eq:58b}) are valid within the full nonlinear theory. They are shown
in \cite{B93} to arise from a specific time-asymmetric component of the
nonlinear gravitational field in $D_e$. The matching of these expressions to
the field of the source in $D_i$ is done in \cite{B95reac}.

The energy, linear momentum and angular momentum which are extracted from
the source by the 1PN radiation reaction force associated with the scalar and
vectorial potentials (\ref{eq:58b}) are to be computed locally (in $D_i$) using
the equations of motion of the source. This computation has been done recently
\cite{B95reac} with a result in perfect agreement with the
corresponding 1PN fluxes of energy and momenta already known from their
computations in the far zone $D_w$ \cite{Th80,BD89}.  Namely,
\begin{eqnarray}
  {dE\over dt} &=& - {G\over c^5} \left\{ {1\over 5} M^{(3)}_{ij}
  M^{(3)}_{ij} + {1\over c^2}\left[{1\over 189} M^{(4)}_{ijk} M^{(4)}_{ijk}
  + {16\over 45} S^{(3)}_{ij} S^{(3)}_{ij} \right] \right\} \nonumber\\
  &&+ O\left( \varepsilon^8 \right)\ ,  \nonumber  \\
  {dP_i\over dt} &=& - {G\over c^7} \left\{ {2\over 63} M^{(4)}_{ijk}
  M^{(3)}_{jk} + {16\over 45} \varepsilon_{ijk} M^{(3)}_{jm} S^{(3)}_{km}
  \right\} + O\left(\varepsilon^9\right)\ , \nonumber \\
  {dS_i\over dt} &=& - {G\over c^5} \varepsilon_{ijk} \left\{ {2\over 5}
  M^{(2)}_{jm} M^{(3)}_{km}+{1\over c^2}\left[{1\over 63} M^{(3)}_{jmn}
  M^{(4)}_{kmn} + {32\over 45} S^{(2)}_{jm} S^{(3)}_{km} \right] \right\}
      \nonumber \\
  &&+ O\left(\varepsilon^8\right)\ , \label{eq:59c}
\end{eqnarray}
where $E$, $P_i$ and $S_i$ denote some local (instantaneous) quantities which
are given by some compact-support integrals over the matter in the source,
and agree to 1PN order with the usual notions of energy, linear momentum
and angular momentum of the source. These formulas show the validity to
1PN order of the (generally postulated) balance equations between the
decreases of energy and momenta in the source and the corresponding
fluxes at infinity. Of particular interest (notably because this is a
purely 1PN effect) is the decrease of the linear momentum $dP_i/dt$,
which corresponds to a net recoil of the center of mass of the source in
reaction to the emission of waves. Numerous authors
\cite{BoR61,PA71,Bek73,Th80} had computed before this
effect as a flux of linear momentum at infinity.

In the case of binary systems, Iyer and Will \cite{IW93,IW95} have obtained,
assuming the validity of the balance equations for energy and angular
momentum, the radiation reaction force in the equations of motion of the
binary to 1PN relative order (which corresponds to the 3.5PN order beyond the
Newtonian acceleration because the reaction effects arise at 2.5PN order).
Their result is valid for a large class of coordinate systems. They also
show \cite{IW95} that the reactive potentials (\ref{eq:58b}) when specialized
to binary systems correspond in their formalism to a unique and consistent
choice of a coordinate system (namely, a ``Burke-Thorne-extended"
coordinate system).  This represents a non-trivial check of the validity
of the reactive potentials (\ref{eq:58b}).

To the next 1.5PN order in the reactive potentials there appears the same
phenomenon
as in the radiation field in $D_w$, namely the occurrence of hereditary effects
associated with tails. This is required if the balance equation for the energy
is to remain valid at this order. That this is indeed the case
has been proved in \cite{BD88} where the dominant hereditary contribution in
the post-Newtonian expansion of the equations of motion of the source (in
$D_i$) was shown to arise at the 4PN order beyond the Newtonian acceleration,
that is precisely at the 1.5PN relative order in the reaction force.
Therefore, one can say that locality in time is lost at the 4PN order (or
order $\varepsilon^8$) in the equations of motion of a self-gravitating
source. This hereditary contribution can be shown in a particular gauge to
modify the scalar reactive potential in (\ref{eq:58b}) by the tail
integral \cite{BD88}
\begin{equation}
  \delta V^{\rm reac} = -{4G^2M\over 5c^8} x_{ij} \int^{+\infty}_0
   dy \left[ \ln \left( {y\over 2b}\right) + {11\over 12} \right]
   M^{(7)}_{ij} (t-y)\ . \label{eq:60}
\end{equation}
This tail contribution in the reaction is recovered in Section III.D of
\cite{B93} using a more systematic method, and the constant
11/12 is computed.

The following picture can be given to this phenomenon.
The waves emitted by the source at all epochs in the past are scattered
onto the spacetime curvature generated by the mass $M$.  The secondary
waves which are produced converge back onto the source at our epoch and
modify its present dynamics. The total reaction force is thus the vectorial
sum of the reaction force due to the present emitted radiation (this is
(\ref{eq:58b})), and of the reaction force due to the incoming tail radiation
(see (\ref{eq:60})). It is shown in Section III.D of \cite{BD92} that the
tail-induced reactive potential $\delta V^{\rm reac}$ is consistent with the
tail integral in (\ref{eq:37'}) in the sense that it gives rise to a term of
order $\varepsilon^8$ in the energy balance equation in (\ref{eq:59c}) which
is in perfect agreement with the contribution of the tail integral
to the flux of energy in $D_w$.

\section{APPLICATION TO INSPIRALLING COMPACT BINARIES} \label{sec:6}

Inspiralling compact binaries are very relativistic (but not fully
relativistic) binary systems of compact objects (neutron stars or black holes)
in their late phase of evolution which precedes immediately the final
coalescence, during which the two objects spiral very rapidly around
each other under the influence of the gravitational radiation reaction forces.

It has been recognized in recent years that a very precise prediction from
general relativity, taking into account many relativistic (or post-Newtonian)
corrections, will be necessary in order that all the potential information
contained in the signals from inspiralling binaries can be deciphered by
the future detectors LIGO and VIRGO
\cite{KSc87}--\cite{KKS95},\cite{BSat94,BSat95}.  The demanded precision
can be calculated using black-hole perturbation techniques in the
special case where the mass of one object is very small as compared with
the other mass \cite{Gal80}--\cite{P95}.  This precision turns out to be
at least 3PN beyond the quadrupole radiation \cite{CFPS93,TNaka94,P95}.  Thus,
inspiralling compact binaries offer us the ideal
application of the previous relativistic formalism ---~and constitute
presently our main motivation for its development.

The information contained in the signals of inspiralling compact binaries
will be extracted from the noisy output of the detector by a technique 
of matched filtering, necessitating the a priori knowledge of the signal.
Crucial to a successful data analysis will be the knowledge of
the instantaneous phase of the binary, namely the angle between the
separation of the two bodies and a fixed direction in the orbital plane.

The time evolution of the orbital phase should in principle be determined from
the gravitational radiation reaction forces acting locally on the orbit.
However, as these forces are only known to 1PN order
\cite{IW93,IW95,B93,B95reac} and 1.5PN order \cite{BD88}, one relies in
practice on an energy balance equation similar to the one in
(\ref{eq:59c}), namely
\begin{equation}
   {dE\over dt} = - {\cal L} \ , \label{eq:61}
\end{equation}
where in both sides one uses some very precise expressions of the (binding)
energy $E$ of the binary and its total luminosity ${\cal L}$ in gravitational
waves (the angular momentum balance equation is not necessary because the
orbit has been circularized by the radiation reaction forces). The fact that
one is obliged presently to {\it postulate} the validity of the energy
balance equation (\ref{eq:61}) to higher order than 1.5PN (for instance
2.5PN as postulated below) is certainly a weak point in the analysis, which
shall have to be improved in future work.

The binding energy $E$ of the binary which appears in the left-hand-side of
(\ref{eq:61}) results from the equations of motion of the binary which have
been obtained to 2.5PN order by Damour and Deruelle \cite{DD81a}--\cite{D83b}
in their study of the binary pulsar PSR~1913+16. For inspiralling compact
binaries one simply needs to specialize these equations to the case
where the orbit is circular (apart from the gradual inspiral).  The
2.5PN--accurate binding energy reads \cite{DD81b}
\begin{equation}
 E = -{c^2\over 2} M\nu x \left\{ 1 - {1\over 12} (9+\nu) x -{1\over 8}
 \left(27-19\nu +{\nu^2\over 3}\right) x^2 + O(x^3) \right\}\ .\label{eq:62}
\end{equation}
(See also the appendix~ B of \cite{B95pn} for a derivation for this formula
from the Damour-Deruelle equations of motion.) The total mass of the binary
is $M=m_1+m_2$, and $\nu=m_1 m_2 /M^2$ represents a particular mass ratio
which is such that $0 < \nu \leq 1/4$, with $\nu = 1/4$ corresponding to
the case of two equal masses and $\nu \to 0$ corresponding to the
test-mass limit for one body. For convenience we denote by
$x=(GM\omega/c^3)^{2/3}$ a small dimensionless post-Newtonian parameter
of order $\varepsilon^2$, where $\omega$ is the instantaneous angular
frequency of the orbit ($\omega=2\pi/P$, where $P$ is the orbital period).

On the other hand, the 2.5PN--accurate luminosity $\cal L$ of the binary is
obtained by application of the formalism above. Namely, one uses the
decomposition
of ${\cal L}$ in terms of observable moments $U_L$ and $V_L$ as given by
(\ref{eq:25}). Then one relates the observable moments to the moments
$M_L$ and $S_L$ by the equations (\ref{eq:36})-(\ref{eq:39}). Actually to
2.5PN order the mass quadrupole moment comprises the dominant tail integral
and the quadrupole-quadrupole interaction terms which include the nonlinear
memory integral, but the tail of tail integral appearing in (\ref{eq:38}) is
negligible. Nonlinear effects are rather easy to compute for binary systems
thanks to the study done in \cite{BS93}, where the required formula is proved
to apply to a binary which is actually spiralling (or decaying). Indeed this
is not obvious a priori because we are considering hereditary effects
which depend on the whole integrated past of the binary (see the appendix B
in \cite{BS93}). In a second stage one must use the expressions of the
moments $M_L$ and $S_L$ as integrals over the source.  These are given
by (\ref{eq:56}) and (\ref{eq:57}) (to 2.5PN order the next-order term
in $M_L$ is a priori also needed, but in fact this term is instantaneous
and does not contribute in the energy loss to this order, see below). A
hard task, carried out to 2PN order in \cite{BDI95} and \cite{WWi95}, is
to reduce all these integrals for compact binary systems.  This entails
using delta functions to describe the compact objects (but this can probably
be justified using the results of \cite{DD81a}--\cite{D83b}). Notably the
main cubically nonlinear term in (\ref{eq:56}) (term with
$P_{ij}\partial^2_{ij}U$) deserves a special treatment. This term has been
computed by two different methods in \cite{BDI95} (see Section
III.C and Appendix B there).  The final result for $\cal L$ reads

\begin{eqnarray}
 {\cal L} &=& {32c^5\over 5G} \nu^2 x^5 \biggl\{ 1
  - \left( {1247\over 336}+{35\over 12}\nu \right) x +4\pi x^{3/2}\nonumber \\
 &&\quad\qquad\qquad\quad + \left(-{44711\over 9072}+{9271\over 504}\nu
   + {65\over 18} \nu^2 \right) x^2 \nonumber \\
 &&\quad\qquad\qquad\quad -\left({8191\over 672}+{535\over 24}\nu\right)
   \pi x^{5/2} + O(x^3) \biggr\} \ . \label{eq:63}
\end{eqnarray}
This expression was already known to Newtonian order by Landau and Lifchitz
\cite{LL0,LL}, and to 1PN order by Wagoner and Will \cite{WagW76} using the
Epstein-Wagoner moments \cite{EW75} (in which infinite surface terms are to
be discarded). It has been re-calculated to 1PN order in \cite{BS89} using the
well-defined compact-support moment (\ref{eq:51}). The 1.5PN order is the
$4\pi$ term which is the contribution of the dominant tail integral present
in (\ref{eq:37'}). The coefficient $4\pi$ is specific to binary systems
moving in circular orbits, and was obtained first by Poisson \cite{P93} using
a black-hole perturbation technique valid in the test-mass limit $\nu \to 0$
(this happens to give the answer, which does not depend on $\nu$, even for
arbitrary mass ratios).
The computation of the tail integral in (\ref{eq:37'}) in the case of binary
systems was done in \cite{BS93,Wi93} with result yielding the $4\pi$ term.
The 2PN term computed in \cite{BDI95} is entirely due to relativistic
corrections in the moments of the source, most importantly the mass
quadrupole moment (\ref{eq:56}) (issued from \cite{B95}). Will and Wiseman
\cite{WWi95} computed independently this term using the Epstein-Wagoner
moments which give the correct result, as proved in \cite{B95}, because
there are no tails at this order (provided that infinite surface terms are
discarded). (But see \cite{WWi95} for improvements of the Epstein-Wagoner
approach, and for how to compute tails within this approach.) The common
result of these 2PN computations was summarized in \cite{BDIWW95}, where it was
also pointed out that the effects of intrinsic rotation of the two objects
are in general negligible as compared to the gravitational effects.
Finally the 2.5PN term has been added in \cite{B95pn} and shown to be due
exclusively to the mass octupole and current quadrupole tail integrals
in (\ref{eq:39}), and to the post-Newtonian corrections in the dominant mass
quadrupole tail integral in (\ref{eq:37'}) or (\ref{eq:38}). The fact that only the
tail integrals give a contribution in ${\cal L}$ to 2.5PN order is special
to the case of a circular orbit \cite{B95pn}. In the non-circular case the
instantaneous terms would also contribute. Incidentally notice that the
nonlinear memory integral is instantaneous in the energy loss, and therefore
gives no contribution in (\ref{eq:63}). Thus the nonlinear memory
effect has rather poor observational consequences for inspiralling binaries.

The tail of tail integral given by the last term in (\ref{eq:38})
gives a contribution to $\cal L$ at the first order which is neglected in
(\ref{eq:63}), namely the order 3PN or $x^3$. This contribution to ${\cal L}$
is computed in \cite{B96tail} (it involves a logarithmic term in the
frequency).  However the whole 3PN term in (\ref{eq:63}) will include
also many contributions coming from the relativistic corrections in the
moments of the source.  These contributions are not under control
presently.  The next 3.5PN order is expected like the 1.5PN and 2.5PN
orders to be due entirely to higher-order multipolar tails, and to
post-Newtonian corrections in low-order multipolar tails. The still higher 4PN
order includes many effects and seems presently difficult to reach within the
post-Newtonian theory.

There is a case, however,  where ${\cal L}$ is known up to 4PN order. This is
the test-mass limit case $\nu \to 0$, which has been computed by
black-hole perturbation techniques.  Tagoshi and Nakamura \cite{TNaka94}
first obtained numerically the coefficients in $\cal L$ up to 4PN order.
Then Sasaki \cite{Sasa94} showed how to solve in an iterative way the
Regge-Wheeler equation for the perturbations of the Schwarzschild black-hole,
and finally, based on this, Tagoshi and Sasaki \cite{TSasa94} obtained
analytically the coefficients in ${\cal L}$ up to 4PN order.  (See also
\cite{SSTT95,Tagoshi95} for the case of a rotating black hole.) The result
(\ref{eq:63}) of
post-Newtonian theory is in perfect agreement in the limit $\nu \to 0$
with the truncation to 2.5PN order of the result of perturbation theory
(tails of tails computed in \cite{B96tail} also perfectly agree). As the
post-Newtonian and perturbation theories are so different, the agreement is
very satisfying, although mandatory of course.

Finally it remains to substitute the 2.5PN expressions of both $E$ and
${\cal L}$ into the energy balance equation (\ref{eq:61}) (assumed to be
true at the 2.5PN level),
in order to obtain the instantaneous phase $\phi$ of the binary as a function
of time. The phase is $\phi = \int \omega dt$ where $\omega$ is the orbital
frequency, from which one deduces $\phi = - \int (\omega/{\cal L}) dE$. The
result is simpler if we use instead of the local time $t$ flowing in the
observer's frame ($t=T-R/c$ with the notation of Section~3) the
adimensional time variable \cite{BDI95}
\begin{equation}
   \Theta = {c^3\nu \over 5GM} (t_c -t)\ , \label{eq:64}
\end{equation}
where $t_c$ denotes the instant of coalescence (at which the frequency goes
formally to infinity).
In terms of this time variable the orbital phase is obtained as
\begin{eqnarray}
\phi (t)&=&\phi_0 - {1\over \nu} \Biggl\{ \Theta^{5/8} + \left(
  {3715\over 8064} + {55\over 96} \nu \right) \Theta^{3/8}
    - {3\pi\over 4} \Theta^{1/4} \nonumber\\
  &&\qquad\qquad + \left( {9275495\over 14450688} + {284875\over 258048} \nu
    + {1855\over 2048} \nu^2  \right) \Theta^{1/8} \nonumber \\
  &&\qquad\qquad - \left( {38645\over 172032} + {15\over 2048} \nu \right)
  \pi \ln \Theta + O (\Theta^{-1/8}) \Biggr\}\ , \label{eq:65}
\end{eqnarray}
where $\phi_0$ is some constant phase determined by initial conditions
(for instance when the signal enters the frequency bandwidth of the detector).
This expression is valid only in the post-Newtonian regime where
$\Theta^{-1} = O(\varepsilon^8)$.

Note that the determination of the phase (\ref{eq:65}) does not constitute
by itself the complete answer to the problem because the waveform
(described by two independent polarization states $h_+$ and $h_\times$)
is also to be computed.  Thus the phase $\phi (t)$, and the orbital
frequency $\omega (t)=\dot\phi (t)$, have to be inserted into the waveform
of the binary, which is itself to be known with the best possible
post-Newtonian precision (though the latter precision does not need to be so
high as in the determination of the phase; see, e.g., \cite{CF94}).
It is known \cite{BS93} that the
waveform introduces in the phase evolution some contributions which add
to those coming purely from the energy balance equation (\ref{eq:61}), but
these extra contributions arise at an order which is higher than 2.5PN.
For the computation of the waveform to 2PN order see \cite{BIWW95}.


\begin{thebibliography}{100}
\bibitem{E16}Einstein A., {\sl Sitzber. Preuss. Akad. Wiss. Berlin} (1916) 688.
\bibitem{E18}Einstein A., {\sl Sitzber. Preuss. Akad. Wiss. Berlin} (1918) 154.
\bibitem{Edd23}Eddington A. S., {\sl The mathematical theory of relativity}
(1923), reprinted by {\sl Cambridge University Press} (1965).
\bibitem{B57}Bondi H., \review Nature, 179, 1957, 1072.
\bibitem{Bo59}Bonnor W. B., \review Philos. Trans. R. Soc. London A, 251,
1959, 233.
\bibitem{BBM62}Bondi H., van der Burg M. G. J., Metzner A. W. K.,
\review Proc. R. Soc. London A, 269, 1962, 21.
\bibitem{S62}Sachs R., \review Proc. R. Soc. London A, 270, 1962, 103.
\bibitem{ADM}Arnowitt R., Deser S., Misner C. W., Gravitation:
an introduction to current research, L. Witten (ed.) (Wiley, New York, 1962).
\bibitem{SY79}Schoen R., Yau S. T., \review Comm. Math. Phys., 65, 1979, 45.
\bibitem{SY82}Schoen R., Yau S. T., \review Phys. Rev. Lett., 48, 1982, 371.
\bibitem{HT75}Hulse R. A., Taylor J. H., \review Astrophys. J. (Letters),
195, 1975, L51.
\bibitem{TFMc79}Taylor J. H., Fowler L. A., Mc Culloch P. M., \review
Nature, 277, 1979, 437.
\bibitem{PeM63}Peters P. C., Mathews J., \review Phys. Rev., 131, 1963, 435.
\bibitem{EH75}Esposito L. W., Harrison E. R., \review Astrophys. J.,
196, 1975, L1.
\bibitem{Wag75}Wagoner R. V., \review Astrophys. J., 196, 1975, L63.
\bibitem{CE70}Chandrasekhar S., Esposito F. P., \review Astrophys. J.,
160, 1970, 153.
\bibitem{C65}Chandrasekhar S., \review Astrophys J., 142, 1965, 1488.
\bibitem{C69}Chandrasekhar S., \review Astrophys J., 158, 1969, 45.
\bibitem{CN69}Chandrasekhar S., Nutku Y., \review Astrophys J., 158, 1969, 55.
\bibitem{Bu69}Burke W. L., unpublished Ph. D. Thesis, California
Institute of Technology (1969).
\bibitem{Th69}Thorne K. S., \review Astrophys. J., 158, 1969, 997.
\bibitem{BuTh70}Burke W. L., Thorne K. S.,  Relativity,
M. Carmeli et al. (eds) (Plenum Press, New York, 1970) p.  208-209.
\bibitem{Bu71}Burke W. L., \review J. Math. Phys., 12, 1971, 401.
\bibitem{AD75}Anderson J. L., DeCanio T. C., \review Gen. Relat. Grav.,
6, 1975, 197.
\bibitem{PaL81}Papapetrou A., Linet B., \review Gen. Relat. Grav., 13, 1981, 335.
\bibitem{Ehl80}Ehlers J., \review Ann. N.Y. Acad. Sci., 336, 1980, 279.
\bibitem{Ker80}Kerlick G. D., \review Gen. Rel. Grav., 12, 1980, 467.
{\sl Gen. Rel. Grav.} {\bf 12} (1980) 521.
\bibitem{BRu81}Breuer R., Rudolph E., \review Gen. Rel. Grav., 13, 1981, 777.
\bibitem{S85}Sch\"afer G., \review Ann. Phys. (N.Y.), 161, 1985, 81.
\bibitem{A87}Anderson J.L., \review Phys. Rev. D, 36, 1987, 2301.
\bibitem{DD81a}Damour T., Deruelle N., \review Phys. Lett., 87A, 1981, 81.
\bibitem{DD81b}Damour T., Deruelle N., \review C. R. Acad. Sc. Paris,
293, 1981, 537.
\bibitem{D82}Damour T., \review C. R. Acad. Sc. Paris, 294, 1982, 1355.
\bibitem{D83a}Damour T., in Ref. \cite{Houches82}, p.~59.
\bibitem{D83b}Damour T., \review Phys. Rev. Lett., 51, 1983, 1019.
\bibitem{LIGO}Abramovici A., Althouse W. E., Drever R. W. P., G\"ursel Y.,
Kawamura S., Raab F. J., Shoemaker D., Siewers L., Spero R. E.,
Thorne K. S., Vogt R. E., Weiss R., Whitcomb S. E., Zucker M. E., \review
Science, 256, 1992, 325.
\bibitem{VIRGO}Bradaschia C., Calloni E., Cobal M.,  Del~Fasbro R.,
Di~Virgilio A., Giazotto A., Holloway L. E., Kautzky H., Michelozzi B.,
Montelatici V., Pascuello D., Velloso W., Gravitation 1990,
Proc. of the Banff Summer Inst., Banff, Alberta, R. Mann and P. Wesson
(eds.), (World Scientific, Singapore, 1991).
\bibitem{EW75}Epstein R., Wagoner R. V., \review Astrophys. J., 197,
1975, 717.
\bibitem{BP60}Bertotti B., Plebanski J., \review Ann. Phys. (N.Y.),
11, 1960, 169.
\bibitem{BoR61}Bonnor W. B., Rotenberg M. A., \review Proc. R. Soc. London
A, 265, 1961, 109.
\bibitem{BoR66}Bonnor W. B., Rotenberg M. A., \review Proc. R. Soc. London
A, 289, 1966, 247.
\bibitem{HR69}Hunter A. J., Rotenberg M. A., \review J. Phys. A, 2, 1969, 34.
\bibitem{Bo74}Bonnor W.  B., Ondes et radiations gravitationnelles
(CNRS, Paris, 1974) p.  73.
\bibitem{ThK75}Thorne K. S., Kov\'acs S. J., \review Astrophys. J.,
200, 1975, 245.
\bibitem{CTh77}Crowley R. J., Thorne K. S., \review Astrophys. J., 215,
1977, 624.
\bibitem{Th80}Thorne K. S., \review Rev. Mod. Phys., 52, 1980, 299.
\bibitem{BD86}Blanchet L., Damour T., \review Philos. Trans. R. Soc.
London A, 320, 1986, 379.
\bibitem{B87}Blanchet L., \review Proc. R. Soc. Lond. A, 409, 1987, 383.
\bibitem{BD88}Blanchet L., Damour T., \review Phys. Rev. D, 37, 1988, 1410.
\bibitem{BD89}Blanchet L., Damour T., \review Ann. Inst. H. Poincar\'e
(Phys. Th\'eorique), 50, 1989, 377.
\bibitem{DI91a}Damour T., Iyer B. R., \review Ann. Inst. H. Poincar\'e
(Phys. Th\'eorique), 54, 1991, 115.
\bibitem{BD92}Blanchet L., Damour T., \review Phys. Rev. D, 46, 1992, 4304.
\bibitem{B93}Blanchet L., \review Phys. Rev. D, 47, 1993, 4392.
\bibitem{B95}Blanchet L., \review Phys. Rev. D, 51, 1995, 2559.
\bibitem{B95reac}Blanchet L., Gravitational radiation reaction to
post-Newtonian order, {\sl Phys.  Rev.  D} (submitted).
\bibitem{B95pn}Blanchet L., Energy losses by gravitational radiation in
inspiralling compact binaries to five halves post-Newtonian order,
{\sl Phys. Rev. D} (in press).
\bibitem{B96tail}Blanchet L., Gravitational wave tails of tails, in preparation.
\bibitem{DI91b}Damour T., Iyer B. R., \review Phys. Rev. D, 43,
1991, 3259.
\bibitem{BS89}Blanchet L., Sch\"afer G., \review Mon. Not. R. Astr. Soc.,
239, 1989, 845.
\bibitem{BS93}Blanchet L., Sch\"afer G., \review Class.  Quantum Grav.,
10, 1993, 2699.
\bibitem{BDI95}Blanchet L., Damour T., Iyer B.  R., \review Phys.  Rev.
D, 51, 1995, 5360.
\bibitem{BDIWW95}Blanchet L., Damour T., Iyer B. R., Will C. M.,
Wiseman A. G., \review Phys. Rev. Lett., 74, 1995, 3515.
\bibitem{BIWW95}Blanchet L., Iyer B. R., Will C. M., Wiseman A. G., \review
Class. Quantum Grav., 13, 1996, 575.
\bibitem{WWi95}Will C. M., Wiseman A. G., in preparation.
\bibitem{Wi92}Wiseman A. G., \review Phys. Rev. D, 46,  1992, 1517.
\bibitem{Wi93}Wiseman A. G., \review Phys. Rev. D, 48, 1993, 4757.
\bibitem{KWW93}Kidder L.  E., Will C.  M., Wiseman A.  G., \review Phys.
Rev.  D, 47,  1993, R4183.
\bibitem{Kidder95}Kidder L. E., \review Phys. Rev. D, 52, 1995, 821.
\bibitem{Houches82}Deruelle N., Piran T. (eds.)
Gravitational Radiation (North Holland Publishing Company, 1983).
\bibitem{300years}Hawking S., Israel W. (eds.), 300
Years of Gravitation (Cambridge University Press, 1987).
\bibitem{Cargese}Carter B., Hartle J. B. (eds.), Gravitation in Astrophysics,
(Plenum Press, 1986).
\bibitem{ICGC91}Iyer B. R., Prasanna A. R., Varma R. K., Vishveshwara C.
V. (eds), {\it Advances in Gravitation and Cosmology}, Wiley (1991).
\bibitem{Th93}Thorne K. S., Proc. of the Fourth Rencontres de
Blois, ed. G.~Fontaine and J. Tran Thanh Van (Editions Fronti\`eres, France,
1993).
\bibitem{Th94}Thorne K. S., Proc. of the 8th Nishinomiya-Yukawa
Symposium on Relativistic Cosmology, M.~Sasaki (ed.) (Universal Acad.
Press, Japan, 1994).
\bibitem{Th94'}Thorne K. S., Proc. of Snowmass 94 Summer
Study on Particle and Nuclear Astrophysics and Cosmology, E.W. Kolb
and R. Peccei (eds.) (World Scientific, Singapore, 1994).
\bibitem{Sc89'}Schutz B. F., \review Class. Quantum Grav., 6, 1989, 1761.
\bibitem{Sc93}Schutz B. F., \review Class. Quantum Grav., 10, 1993, S135.
\bibitem{BonaM94}Bonazzola S., Marck J. A., \review Ann. Rev. Nucl.
Part. Sci., 45, 1994, 655.
\bibitem{W94}Will C. M., Proc. of the 8th Nishinomiya-Yukawa
Symposium on Relativistic Cosmology, M.~Sasaki (ed.) (Universal Acad.
Press, Japan, 1994).
\bibitem{LL0}Landau L. D., Lifshitz E. M., Teoriya Polya (Nauka,
Moscow, 1941).
\bibitem{LL}Landau L. D., Lifshitz E. M., The classical theory
of fields (Pergamon, Oxford, 1971).
\bibitem{Fock}Fock V. A., Theory of Space, Time and
Gravitation (Pergamon, London 1959).
\bibitem{MTW}Misner C. W., Thorne K. S., Wheeler J. A.,
Gravitation (Freeman, San Francisco, 1973).
\bibitem{Mi74}Miller B. D., \review Astrophys. J., 187, 1974, 609.
\bibitem{S83}Sch\"afer G., \review Lett. Nuovo Cim., 36, 1983, 105.
\bibitem{Lagerst}Lagerstr\"om P. A., Howard L.N., Liu C. S.,
Fluid mechanics and singular perturbations; a collection  of papers
by Saul Kaplun (Academic, New York, 1967).
\bibitem{DS90}Damour T., Schmidt B., \review J. Math. Phys., 31,
1990, 2441.
\bibitem{G70}Geroch R., \review J. Math. Phys., 11, 1970, 2580.
\bibitem{H74}Hansen R. O., \review J. Math. Phys., 15, 1974, 46.
\bibitem{Gu83}G\"ursel Y., \review Gen. Rel. Grav., 15, 1983, 737.
\bibitem{A79}Anderson J.L., Isolated Gravitating Systems in General
Relativity, J. Ehlers (ed) (North-Holland, 1979) p. 289.
\bibitem{AKKM82}Anderson J.L., Kates R.E., Kegeles L.S., Madonna
R.G., \review Phys. Rev. D, 25, 1982, 2038.
\bibitem{FuSc83}Futamase T., Schutz B.F., \review Phys. Rev. D, 28,
1983, 2363.
\bibitem{Fu83}Futamase T., \review Phys. Rev. D, 28, 1983, 2373.
\bibitem{P63}Penrose R., \review Phys. Rev. Lett., 10, 1963, 66.
\bibitem{P65}Penrose R., \review Proc. R. Soc. Lond. A, 284, 1965, 159.
\bibitem{G77}Geroch R., Asymptotic structure of space-time,
P. Esposito and L. Witten (eds.) (Plenum Press, New York, 1977).
\bibitem{Sch79}Schmidt B. G., Isolated gravitating systems in
general relativity, J. Ehlers (ed.) (North-Holland, Amsterdam, 1979).
\bibitem{A84}Ashtekar A., General relativity and gravitation
(proceedings of GR10), B. Bertotti et al (eds.) (Reidel, Dordrecht, 1984).
\bibitem{Mad}Madore J., \review Ann. Inst. H. Poincar\'e, 12,
1970, 285~and~365.
\bibitem{GH78}Geroch R., Horowitz G. T., \review Phys. Rev. Lett., 40,
1978, 203.
\bibitem{BaPr73}Bardeen J. M., Press W. H., \review J. Math. Phys.,
14, 1973, 7.
\bibitem{WalW79}Walker M., Will C. M., \review Phys. Rev. D, 19, 1979, 3495.
\bibitem{D86'}Damour T., Proc. of the Fourth Marcel Grossmann
Meeting on General Relativity, R. Ruffini (ed.) (Elsevier Science Publishers,
1986), p. 365.
\bibitem{YCB1}Four\`es-Bruhat Y., \review Acta Math., 88, 1952, 141.
\bibitem{CTJNew68}Couch W. E., Torrence R. J., Janis A. I.,
 Newman E. T., \review J. Math.  Phys., 9, 1968, 484.
\bibitem{KNew68}Kundt W., Newman E. T., \review J. Math. Phys., 9, 1968, 2193.
\bibitem{McLen86}Carminati J.,  McLenaghan R. G., \review Ann. Inst.
Henri Poincar\'e (Physique Th\'eorique) A, 44, 1986, 115.
\bibitem{YCB90}Choquet-Bruhat Y., McLenaghan R., \review C. R. Acad.
Sci. Paris, 311, 1990, 483.
\bibitem{CH71}Couch W. E., Hallidy W. H., \review J. Math. Phys., 12,
1971, 2170.
\bibitem{S90}Sch\"afer G., \review Astron. Nachr., 311, 1990, 213.
\bibitem{BSat94}Blanchet L., Sathyaprakash B. S., \review Class. Quantum Grav.,
11, 1994, 2807.
\bibitem{BSat95}Blanchet L., Sathyaprakash B. S.,
\review Phys. Rev. Letters, 74, 1995, 1067.
\bibitem{P83}Payne P. N., \review Phys. Rev. D, 28, 1983, 1894.
\bibitem{B90}Blanchet L., th\`ese d'habilitation, Universit\'e P.
et M.  Curie (1990) (unpublished).
\bibitem{Chr91}Christodoulou D., \review Phys. Rev. Lett., 67, 1991, 1486.
\bibitem{Th92}Thorne K. S., \review Phys. Rev. D, 45, 1992, 520.
\bibitem{BTh87}Braginsky V. B., Thorne K. S., \review Nature, 327,
1987, 123.
\bibitem{WiW91}Wiseman A. G., Will C. M., \review Phys. Rev. D, 44,
1991, R2945.
\bibitem{CMM77}Campbell W. B., Macek J., Morgan T. A., \review Phys.
Rev. D, 15, 1977, 2156.
\bibitem{BD84}Blanchet L., Damour T., \review Phys. Lett., 104A, 1984, 82.
\bibitem{PA71}Papapetrou A., \review Ann. Inst. H. Poincar\'e, XIV, 1971, 79.
\bibitem{Bek73}Bekenstein J. D., \review Astrophys. J., 183, 1973, 657.
\bibitem{IW93}Iyer B. R., Will C. M., \review Phys. Rev. Lett., 70,
1993, 113.
\bibitem{IW95}Iyer B. R., Will C. M., \review Phys. Rev. D, 52, 1995, 6882.
\bibitem{KSc87}Kr\'olak A., Schutz B. F., \review Gen. Rel. Grav., 19, 1987,
1163.
\bibitem{K89}Kr\'olak A., Gravitational Wave Data Analysis, B.F. Schutz
(ed) (Kluwer Academic Publishers, 1989).
\bibitem{LW90}Lincoln C. W., Will C. M., \review Phys. Rev. D, 42, 1990, 1123.
\bibitem{3mn}Cutler C., Apostolatos T. A., Bildsten L., Finn L. S.,
Flanagan E. E., Kennefick D., Markovic D. M., Ori A., Poisson E., Sussman G. J.,
Thorne K. S., \review Phys. Rev. Lett., 70, 1993, 2984.
\bibitem{FCh93}Finn L. S., Chernoff D. F., \review Phys. Rev. D, 47, 1993, 2198.
\bibitem{CF94}Cutler C., Flanagan E., \review Phys. Rev. D, 49, 1994, 2658.
\bibitem{PW95}Poisson E., Will C. M., \review Phys. Rev. D, 52, 1995, 848.
\bibitem{KKS95}Kr\'olak A., Kokkotas K. D., Sch\"afer G., \review Phys.
Rev. D, 52, 1995, 2089.
\bibitem{Gal80}Gal'tsov D. V., Matiukhin A. A., Petukhov V. I.,
\review Phys. Lett., 77A, 1980, 387.
\bibitem{P93}Poisson E., \review Phys. Rev. D, 47, 1993, 1497.
\bibitem{CFPS93}Cutler C., Finn L. S., Poisson E., Sussmann G.J.,
\review Phys. Rev. D, 47, 1993, 1511.
\bibitem{TNaka94}Tagoshi H., Nakamura T., \review Phys. Rev. D,
49, 1994, 4016.
\bibitem{Sasa94}Sasaki M., \review Prog. Theor. Phys., 92, 1994, 17.
\bibitem{TSasa94}Tagoshi H., Sasaki M., \review Prog. Theor. Phys.,
92, 1994, 745.
\bibitem{P95}Poisson E., \review Phys. Rev. D, 52, 1995, 5719.
\bibitem{WagW76}Wagoner R. V., Will C. M., \review Astrophys. J., 210, 1976,
764.
\bibitem{SSTT95}Shibata M., Sasaki M., Tagoshi H., Tanaka T., \review
Phys. Rev. D, 51, 1995, 1646.
\bibitem{Tagoshi95}Tagoshi H., \review Prog. Theor. Phys., 93, 1995, 307.
\end{thebibliography}
\end{document}